\documentclass[10pt]{article}

\usepackage[english]{babel}
\usepackage{amsmath}
\usepackage{amssymb}
\usepackage{bm}
\usepackage{bbm}
\usepackage{mathrsfs,color}
\usepackage{graphics,graphicx,theorem}
\usepackage{dsfont}
\usepackage{setspace}

\usepackage{natbib}
\usepackage{multirow}
\usepackage{hhline}
\usepackage{url}
\usepackage{hyperref}

\hypersetup{
	colorlinks=true,
	urlcolor=blue,
	citecolor=blue,
	linktoc=all,
	linkcolor=red} 

\usepackage[scriptsize]{subfigure}
\allowdisplaybreaks

\makeatletter
\renewcommand\@biblabel[1]{}
\makeatother

\def\bSig\mathbf{\Sigma}

\newcommand{\E}{\mathbb{E}}

\renewcommand{\P}{\mathbb{P}}

\newcommand{\ddr}{\mathrm{d}}

\newcommand{\edr}{\mathrm{e}}

\newcommand{\qed}{$\square$}

\newcommand{\smallon}{o\left(\frac{1}{n}\right)}

\newcommand{\argmax}{\mathop{\arg\max}}
\newcommand{\Vnk}{V_{h,(n,k_{n})}}
\newcommand{\Vnuk}{V_{h,(n+1,k_{n})}}
\newcommand{\Vnuku}{V_{h,(n+1,k_{n}+1)}}

\newcommand{\Vgnk}{V_{g,(n,k_{n})}}
\newcommand{\Vgnuk}{V_{g,(n+1,k_{n})}}

\def\simiid{\stackrel{\mbox{\scriptsize{iid}}}{\sim}}

\usepackage[top=1.2in,bottom=1.2in,left=1.2in,right=1.2in]{geometry}

\begin{document}

\title{\bf {\Large{Bayesian nonparametric inference for discovery probabilities:\\credible intervals and large sample asymptotics}}}
\author{
Julyan Arbel\\
\normalsize{Bocconi University, Milan and Collegio Carlo Alberto, Italy}\\
\normalsize{email: \href{mailto:julyan.arbel@unibocconi.it}{julyan.arbel@unibocconi.it}}
\medskip\\
Stefano Favaro\\
\normalsize{University of Torino and Collegio Carlo Alberto, Italy}\\
\normalsize{email: \href{mailto:stefano.favaro@unito.it}{stefano.favaro@unito.it}}
\medskip\\
Bernardo Nipoti\\
\normalsize{Trinity College Dublin, Ireland}\\
\normalsize{email: \href{mailto:nipotib@tcd.ie}{nipotib@tcd.ie}}
\medskip\\
Yee Whye Teh\\
\normalsize{University of Oxford, UK}\\
\normalsize{email: \href{mailto:y.w.teh@stats.ox.ac.uk}{y.w.teh@stats.ox.ac.uk}}
}
\date{}
\maketitle
\thispagestyle{empty}

\setcounter{page}{1}
\begin{center}
\textbf{Abstract} 
\end{center}
Given a sample of size $n$ from a population of individuals belonging to different species with unknown proportions, a popular problem of practical interest consists in making inference on the probability $D_{n}(l)$ that the $(n+1)$-th draw coincides with a species with frequency $l$ in the sample, for any $l=0,1,\ldots,n$. This paper contributes to the methodology of Bayesian nonparametric inference for $D_{n}(l)$. Specifically, under the general framework of Gibbs-type priors we show how to derive credible intervals for a  Bayesian nonparametric estimation of $D_{n}(l)$, and we investigate the large $n$ asymptotic behaviour of such an estimator. Of particular interest are special cases of our results obtained under the specification of the two parameter Poisson--Dirichlet prior and the normalized generalized Gamma prior, which are two of the most commonly used Gibbs-type priors. With respect to these two prior specifications, the proposed results are illustrated through a simulation study and a benchmark Expressed Sequence Tags dataset. To the best our knowledge, this illustration provides the first comparative study between the two parameter Poisson--Dirichlet prior and the normalized generalized Gamma prior in the context of Bayesian nonparemetric inference for $D_{n}(l)$.
\vspace*{.1in}

\noindent\textsc{Keywords}: {Asymptotics; Bayesian nonparametrics; credible intervals; discovery probability; Gibbs-type priors; Good--Turing estimator; normalized generalized Gamma prior; smoothing technique; two parameter Poisson--Dirichlet.} 

\def\thefigure{\arabic{figure}}
\def\thetable{\arabic{table}}

\fontsize{10.95}{14pt plus.8pt minus .6pt}\selectfont

\setcounter{equation}{0}
\section{Introduction}
\label{s:intro}

The problem of estimating discovery probabilities arises when an experimenter is sampling from a population of individuals $(X_{i})_{i\geq1}$ belonging to an (ideally) infinite number of species $(Y_{i})_{i\geq1}$ with unknown proportions $(q_{i})_{i\geq1}$. Given an observable sample $\boldsymbol{X}_{n}=(X_{1},\ldots,X_{n})$, interest lies in estimating the probability that the $(n+1)$-th draw coincides with a species with frequency $l$ in $\boldsymbol{X}_{n}$, for any $l=0,1,\ldots,n$. This probability is denoted by $D_{n}(l)$ and referred to as the $l$-discovery, while \emph{discovery probabilities} is used to address this class of probabilities. In terms of the species proportions $q_{i}$'s, we can write
\begin{equation}\label{ldisc}
D_{n}(l)=\sum_{i\geq1}q_{i}\mathds{1}_{\{l\}}(\tilde N_{i,n}),
\end{equation}
where $\tilde N_{i,n}$ denotes the frequency of the species $Y_{i}$ in the sample. Here $D_{n}(0)$ is the proportion of yet unobserved species or, equivalently, the probability of discovering a new species. The reader is referred to \citet{Bun(93)} and \citet{Bun(14)} for  comprehensive reviews on the full range of statistical approaches, parametric and nonparametric, as well as frequentist and Bayesian, for estimating the $l$-discovery and related quantities. 
The term discovery probability is also used in the literature to refer to a more general class of probabilities that originate when considering an additional unobserved sample of size $m\geq 0$. For instance, in this framework and conditionally on $\boldsymbol{X}_n$, \citet{Lij(07)} consider the problem of estimating the probability that $X_{n+m+1}$ is new, while \citet{Fav(12)} focus on the so-called $m$-step $l$-discovery,  the probability that $X_{n+m+1}$ coincides with a species that has been observed with frequency $l$ in the enlarged sample of size $n+m$. According to this terminology, the discovery probability $D_n(l)$ introduced in \eqref{ldisc} is the 0-step $l$-discovery.

The estimation of the $l$-discovery has found numerous applications in ecology and linguistics, and its importance has grown considerably in recent years, driven by challenging applications in bioinformatics, genetics, machine learning, design of experiments, etc. For examples, \citet{Efr(76)} and \citet{Chu(91)} discuss applications in empirical linguistics; \citet{Goo(53)} and \citet{Cha(92)}, among many others, discuss the probability of discovering new species of animals in a population; \citet{Mao(02)}, \citet{Nav(08)}, \citet{Lij(07a)}, and \citet{Gui(14)} study applications in genomics and molecular biology; \citet{Zha(05)} considers applications to network species sampling problems and data confidentiality; \citet{Car(15)} discuss applications arising from bipartite and sparse random graphs; \citet{Ras(79)} and \citet{Cha(09)} investigate optimal stopping procedures in finding new species; \citet{Bub(13)} study applications within the framework of multi-armed bandits for security analysis of electric power systems.  

This paper contributes to the methodology of Bayesian nonparametric inference for $D_{n}(l)$. As observed in \citet{Lij(07)} for the discovery probability of new species ($0$-discovery $D_{n}(0)$), a natural Bayesian nonparametric approach for estimating $D_{n}(l)$ consists in randomizing the $q_{i}$'s. Specifically, consider the random probability measure $Q=\sum_{i\geq1}q_{i}\delta_{Y_{i}}$, where $(q_{i})_{i\geq1}$ are nonnegative random weights such that $\sum_{i\geq1}q_{i}=1$ almost surely, and $(Y_{i})_{i\geq1}$ are random locations independent of $(q_{i})_{i\geq1}$ and independent and identically distributed as a nonatomic probability measure $\nu_{0}$ on a space $\mathbb{X}$. Then, it is assumed that
\begin{equation}\label{eq:exchangeable_model}
\begin{split}
X_i\,|\,Q & \quad\simiid\quad Q, \qquad i=1,\ldots,n\\
 Q & \quad\sim\quad \mathscr{Q},
 \end{split}
\end{equation}
for any $n\geq1$, where $\mathscr{Q}$ is the prior distribution over the species composition. Under the Bayesian nonparametric model \eqref{eq:exchangeable_model}, the estimator of $D_{n}(l)$ with respect to a squared loss function, say $\hat{\mathcal{D}}_{n}(l)$, arises from the predictive distributions characterizing $(X_{i})_{i\geq1}$. Specifying $Q$ in the large class of Gibbs-type random probability measures by \citet{Pit(03)}, we consider the problem of deriving credible intervals for $\hat{\mathcal{D}}_{n}(l)$, and study the large $n$ asymptotic behaviour of $\hat{\mathcal{D}}_{n}(l)$. Before introducing our results, we review some aspects of $\hat{\mathcal{D}}_{n}(l)$.\\

\subsection{Preliminaries on $\hat{\mathcal{D}}_{n}(l)$}\label{s:pre}

Let $\boldsymbol{X}_{n}$ be a sample from a Gibbs-type random probability measure $Q$, featuring $K_{n}=k_{n}$ species $X_{1}^{\ast},\ldots,X_{K_{n}}^{\ast}$, the unique values of $\boldsymbol{X}_{n}$ recorded in order of appearance, with corresponding frequencies $(N_{1,n},\ldots,N_{K_{n},n})=(n_{1,n},\ldots,n_{k_{n},n})$. Here for every $i=1,2,\ldots,k_n$, there exists a non-negative integer $\xi_i$ such that $X_{i}^{\ast}=Y_{\xi_i}$ and $N_{i,n} = \tilde N_{\xi_i,n}$, where $(Y_{n})_{n\geq 1}$ is the sequence of random atoms in the definition of $Q$.
Let $\sigma\in(0,1)$ and $(V_{n,k})_{k\leq n,n\geq1}$ be a triangular array of nonnegative weights such that $V_{1,1}=1$ and $V_{n,k}=(n-\sigma k)V_{n+1,k}+V_{n+1,k+1}$. According to de Finetti's representation theorem, $\boldsymbol{X}_{n}$ is part of an exchangeable sequence $(X_{i})_{i\geq1}$ whose distribution has been characterized in \citet{Pit(03)} and \citet{Gne(06)} as follows: for any set $A$ in the Borel sigma-algebra of $\mathbb{X}$,
\begin{equation}\label{eq:gibbs_pred}
\P[X_{n+1}\in A\,|\,\boldsymbol{X}_{n}]=\frac{V_{n+1,k_{n}+1}}{V_{n,k_{n}}}\nu_{0}(A)+\frac{V_{n+1,k_{n}}}{V_{n,k_{n}}}\sum_{i=1}^{k_{n}}(n_{i,n}-\sigma)\delta_{X_{i}^{\ast}}(A).
\end{equation}
The conditional probability \eqref{eq:gibbs_pred} is referred to as the predictive distribution of $Q$. Two peculiar features of $Q$ emerge directly from \eqref{eq:gibbs_pred}: the probability that $X_{n+1}\notin\{X_{1}^{\ast},\ldots,X_{K_{n}}^{\ast}\}$ depends only on $k_{n}$; the probability that $X_{n+1}=X_{i}^{\ast}$ depends only on $(k_{n},n_{i,n})$. See \citet{Deb(15)} for a review on Gibbs-type priors in Bayesian nonparametrics.

Two of the most commonly used nonparametric priors are of Gibbs-type; the two-parameter Poisson--Dirichlet (PD) prior in \citet{Pit(95)} and \citet{Pit(97)}; the normalized generalized Gamma (GG) prior in \citet{Pit(03)} and \citet{Lij(07b)} (see also \citet{prunster2002random},\citet{Jam(02)},\citet{Lij(03)}, and \citet{regazzini2003distributional} for early appearance of normalized GG). The Dirichlet process of \citet{Fer(73)} {can be recovered from both priors by letting} $\sigma\rightarrow0$. For any $\sigma\in(0,1)$, $\theta>-\sigma$ and $\tau>0$, the predictive distributions of the two-parameter PD and the normalized GG priors are of the form \eqref{eq:gibbs_pred} where $V_{n,k_{n}}$, respectively, are
\begin{equation}\label{eq:pred_pd}
\frac{\prod_{i=0}^{k_{n}-1}(\theta+i\sigma)}{(\theta)_{n}} \quad \text{and} \quad
\frac{\sigma^{k_{n}-1}\text{e}^{\tau^{\sigma}}}{\Gamma(n)}\sum_{i=0}^{n-1}{n-1\choose i}(-\tau)^{i}\Gamma\left(k_{n}-\frac{i}{\sigma};\tau^{\sigma}\right),
\end{equation}
where $(a)_{n}:=\prod_{0\leq i\leq n-1}(a+i)$ with $(a)_{0}:=1$, and $\Gamma(a,b):=\int_{b}^{+\infty}x^{a-1}\exp\{-x\}\ddr x$. See \citet{Pit(95),Lij(07b)} for details on \eqref{eq:pred_pd}. According to \eqref{eq:gibbs_pred}, the parameter $\sigma$ admits an interpretation in terms of the distribution of $K_{n}$: the larger $\sigma$, the higher is the number of species and, among these, most of them have small abundances. In other terms, the larger the $\sigma$ the flatter is the distribution of $K_{n}$. The parameters $\theta$ and $\tau$ are location parameters, the bigger they are the larger the expected number of species tends to be.

Denote by $M_{l,n}$ the number of species with frequency $l$ in $\boldsymbol{X}_{n}$, and by $m_{l,n}$ the corresponding observed value. An estimator $\hat{\mathcal{D}}_{n}(l)$ arises from \eqref{eq:gibbs_pred} by suitably specifying the Borel set $A$. In particular, if $A_{0}:=\mathbb{X}\setminus\{X_{1}^{\ast},\ldots,X_{K_{n}}^{\ast}\}$ and $A_{l}:=\{X_{i}^{\ast}:N_{i,n}=l\}$, for any $l=1,\ldots,n$, then one has
\begin{gather}\label{eq:disc_new}
\hat{\mathcal{D}}_{n}(0)=\P[X_{n+1}\in A_{0}\,|\,\boldsymbol{X}_{n}]=\E[Q(A_{0})\,|\,\boldsymbol{X}_{n}]=\frac{V_{n+1,k_{n}+1}}{V_{n,k_{n}}},\\
\label{eq:disc_old}
\hat{\mathcal{D}}_{n}(l)=\P[X_{n+1}\in A_{l}\,|\,\boldsymbol{X}_{n}]=\E[Q(A_{l})\,|\,\boldsymbol{X}_{n}]=(l-\sigma)m_{l,n}\frac{V_{n+1,k_{n}}}{V_{n,k_{n}}}.
\end{gather}
Estimators \eqref{eq:disc_new} and \eqref{eq:disc_old} provide Bayesian counterparts to the celebrated Good--Turing estimator $\check{\mathcal{D}}_{n}(l)=(l+1)m_{l+1,n}/n$, for any $l=0,1,\ldots,n-1$, which is a frequentist nonparametric estimator of $D_{n}(l)$ introduced in \citet{Goo(53)}. The most notable difference between $\hat{\mathcal{D}}_{n}(l)$ and $\check{\mathcal{D}}_{n}(l)$ consists in the use of the information in $\boldsymbol{X}_{n}$: $\check{\mathcal{D}}_{n}(l)$ is a function of $m_{l+1,n}$, and not of $(k_{n},m_{l,n})$ as one would intuitively expect for an estimator of $D_{n}(l)$. See \citet{Fav(12)} for details. 

Under the two-parameter PD prior, \citet{Fav(15)} established a large $n$ asymptotic relationship between $\hat{\mathcal{D}}_{n}(l)$ and $\check{\mathcal{D}}_{n}(l)$. Due to the irregular behaviour of the $m_{l,m}$'s, the peculiar dependency on $m_{l+1,n}$ makes $\check{\mathcal{D}}_{n}(l)$ a sensible estimator only if $l$ is sufficiently small with respect to $n$. See for instance \citet{Goo(53)} and \citet{Sam(01)} for examples of absurd estimates determined by $\check{\mathcal{D}}_{n}(l)$. In order to overcome this drawback, \citet{Goo(53)} suggested smoothing $(m_{l,n})_{l\geq1}$ to a more regular series $(m_{l,n}^{\prime})_{l\geq1}$, where $m_{l,n}^{\prime}=p_{l}k_{n}$ with $\mathscr{S}=(p_{l})_{l\geq1}$ being nonnegative weights such that $\sum_{l\geq0}(l+1)m_{l+1,n}^{\prime}/n=1$. The resulting smoothed estimator is
\begin{displaymath}
\check{\mathcal{D}}_{n}(l;\mathscr{S})=(l+1)\frac{m^{\prime}_{l+1,n}}{n}.
\end{displaymath}
See Chapter 7 in \citet{Sam(01)} and references therein for a comprehensive account on smoothing techniques for $\check{\mathcal{D}}_{n}(l)$. According to Theorem 1 in \citet{Fav(15)}, as $n$ becomes large, $\hat{\mathcal{D}}_{n}(l)$ is asymptotically equivalent to $\check{\mathcal{D}}_{n}(l;\mathscr{S}_{\text{PD}})$, where $\mathscr{S}_{\text{PD}}$ denotes a smoothing rule such that
\begin {equation}\label{eq:smoother}
m_{l,n}^{\prime}=\frac{\sigma(1-\sigma)_{l-1}}{l!}k_{n}.
\end{equation}
While the smoothing approach was introduced as an ad hoc tool for post processing the irregular $m_{l,n}$'s in order to improve the performance of $\check{\mathcal{D}}_{n}(l)$, Theorem 1 in \citet{Fav(15)} shows that, for a large sample size $n$, a similar smoothing mechanism underlies the Bayesian nonparametric framework \eqref{eq:exchangeable_model} with a two-parameter PD prior. Interestingly, the smoothing rule $\mathscr{S}_{\text{PD}}$ has been proved to be a generalization of the Poisson smoothing rule discussed in \citet{Goo(53)} and \citet{Eng(78)}.

\subsection{Contributions of the paper and outline}

The problem of associating a measure of uncertainty to Bayesian nonparametric estimators for discovery probabilities was first addressed in \citet{Lij(07)} where estimates of the probability of observing a new species are endowed with highest posterior density intervals. \citet{Fav(15)} derive asymptotic posterior credible intervals covering also the case of species already observed with a given frequency. These contributions ultimately rely on the presence of an additional unobserved sample. While the approach of \citet{Lij(07)} cannot be used to associate a measure of uncertainty to $\hat{\mathcal{D}}_n(0)$, where such additional sample is not considered, the approach of \cite{Fav(15)} could be taken to derive approximate credible intervals for $\hat{\mathcal{D}}_n(l)$, $l=0,1,\ldots,n$. Nonetheless, due to the asymptotic nature of the approach, the resulting credible intervals are likely to perform poorly for moderate sample size $n$ by underestimating the uncertainty associated to the estimators. They then leave essentially unaddressed the issue of quantifying the uncertainty associated to the estimators $\hat{\mathcal{D}}_n(l)$, for $l=0,1,\ldots,n$. 
In this paper we provide an answer to this problem. With a slight abuse of notation, throughout the paper we write $X\,|\,Y$ to denote a random variable whose distribution coincides with the conditional distribution of $X$ given $Y$. Since $\hat{\mathcal{D}}_{n}(l)=\E[Q(A_{l})\,|\,\boldsymbol{X}_{n}]$, the problem of deriving credible intervals for $\hat{\mathcal{D}}_{n}(l)$ boils down to the problem of characterizing the distribution of $Q(A_{l})\,|\,\boldsymbol{X}_{n}$, for any $l=0,1,\ldots,n$. Indeed this distribution takes on the interpretation of the posterior distribution of $D_{n}(l)$ with respect to the sample $\boldsymbol{X}_{n}$. For any Gibbs-type priors we provide an explicit expression for $\mathcal{E}_{n,r}(l):=\E[(Q(A_{l}))^{r}\,|\,\boldsymbol{X}_{n}]$, for any $r\geq1$. Due to the bounded support of $Q(A_{l})\,|\,\boldsymbol{X}_{n}$, the sequence $(\mathcal{E}_{n,r}(l))_{r\geq1}$ characterizes uniquely the distribution of $Q(A_{l})\,|\,\boldsymbol{X}_{n}$ and, in principle, it can be used to obtain an approximate evaluation of such a distribution. In particular, under the two-parameter PD prior and the normalized GG prior we present an explicit and simple characterization of the distribution of $Q(A_{l})\,|\,\boldsymbol{X}_{n}$.

We also study the large $n$ asymptotic behaviour of $\hat{\mathcal{D}}_{n}(l)$, thus extending Theorem 1 in \citet{Fav(15)} to Gibbs-type priors. Specifically, we show that,  as $n$ tends to infinity, $\hat{\mathcal{D}}_{n}(0)$ and $\hat{\mathcal{D}}_{n}(l)$ are asymptotically equivalent to $\hat{\mathcal{D}}_{n}^{\prime}(0)=\sigma k_{n}/n$ and $\hat{\mathcal{D}}^{\prime}_{n}(l)=(l-\sigma)m_{l,n}/n$, respectively. In other terms, at the order of asymptotic equivalence, any Gibbs-type prior leads to the same approximating estimator $\hat{\mathcal{D}}_{n}^{\prime}(l)$. As a corollary we obtain that $\hat{\mathcal{D}}_{n}(l)$ is asymptotically equivalent to the smoothed Good--Turing estimator $\check{\mathcal{D}}_{n}(l;\mathscr{S}_{\text{PD}})$, namely $\mathscr{S}_{\text{PD}}$ is invariant with respect to any Gibbs-type prior. Refinements of $\hat{\mathcal{D}}^{\prime}_{n}(l)$ are presented for the two-parameter PD prior and the normalized GG prior. A thorough study of the large $n$ asymptotic behaviour of \eqref{eq:gibbs_pred} reveals that  for $V_{n,k_{n}}$ in \eqref{eq:pred_pd}  the estimator $\hat{\mathcal{D}}_{n}(l)$ admits large $n$ asymptotic expansions whose first order truncations coincide with $\hat{\mathcal{D}}^{\prime}_{n}(l)$, and that  second order truncations depend on $\theta>-\sigma$ and $\tau>0$, respectively, thus providing approximating estimators that differ. A discussion of these second order asymptotic refinements is presented with a view towards the problem of finding corresponding refinements of the relationship between $\hat{\mathcal{D}}_{n}(l)$ and $\check{\mathcal{D}}_{n}(l;\mathscr{S}_{\text{PD}})$. 

The estimators $\hat{\mathcal{D}}_{n}(l)$ depend on the values assigned to the involved parameters (see e.g. the sensitivity analysis in \citep{Fav(15)}  for the two-parameter PD case) that therefore must be suitably estimated, e.g. via an empirical Bayes approach. Taking into account the method used to estimate the parameters characterizing the underlying Gibbs-type prior would then make the analysis of the asymptotic behaviour of $\hat{\mathcal{D}}_{n}(l)$ more thorough, but we consider the parameters as fixed. We want to stick to the original Bayesian nonparametric framework for the estimation of discovery probabilities, as set forth in \citet{Lij(07)}, and we believe that this best serves the purpose of comparing the asymptotic behaviour {of the two} classes of estimators, highlighting the effect of the parameters in both.

Our results are illustrated in a simulation study and in the analysis of a benchmark dataset of Expressed Sequence Tags (ESTs), which are short cDNA sub-sequences highly relevant for gene identification in organisms \citep[see][]{Lij(07a)}. To the best of our knowledge, only the two-parameter PD prior has been so far applied in the context of Bayesian nonparametric inference for the discovery probability. We consider the two-parameter PD prior and the normalized GG prior. It turns out that  the two-parameter PD prior leads to estimates of the $l$-discovery, as well as associated credible intervals, that are close to those obtained under the normalized GG prior specification. This surfaces due to a representation of the two-parameter PD prior in terms of a suitable mixture of normalized GG priors. Credible intervals for $\hat{\mathcal{D}}_{n}(l)$ are also compared with corresponding confidence intervals for the Good--Turing estimator, which as obtained by \citet{Mao(04)} and \citet{Baa(01)}. A second numerical illustration is devoted to the large $n$ asymptotic behaviour of $\hat{\mathcal{D}}_{n}(l)$, by using simulated data we compare the exact estimator $\hat{\mathcal{D}}_{n}(l)$ with its first order and second order approximations.

In Section 2 we present some distributional results for $Q(A_{l})\,|\,\boldsymbol{X}_{n}$; these results provide a fundamental tool for deriving credible intervals for the Bayesian nonparametric estimator $\hat{\mathcal{D}}_{n}(l)$. In Section 3 we investigate the large $n$ asymptotic behaviour of $\hat{\mathcal{D}}_{n}(l)$, and we discuss its relationship with smoothed Good--Turing estimators. Section 4 contains some numerical illustrations. Proofs, technical derivations and additional illustrations are available in the Appendix.


\setcounter{equation}{0} 
\section{Credible intervals for $\hat{\mathcal{D}}_{n}(l)$}\label{s:ci}

An integral representation for the $V_{n,k_{n}}$'s characterizing the predictive distributions \eqref{eq:gibbs_pred} was introduced by \citet{Pit(03)},  and leads to a useful parameterization for Gibbs-type priors. See also \citet{Gne(06)} for details. For any $\sigma\in(0,1)$ let $f_{\sigma}$ be the density function of a positive $\sigma$-stable random variable,  $\int_{0}^{+\infty}\exp\{-tx\}f_{\sigma}(x)\ddr x=\exp\{-t^{\sigma}\}$ for any $t>0$. Then, for some nonnegative function $h$, one has
\begin{equation}\label{eq_weight}
V_{n,k_{n}}=V_{h,(n,k_{n})}:=\frac{\sigma^{k_{n}}}{\Gamma(n-\sigma k_{n})}\int_{0}^{+\infty}h(t)t^{-\sigma k_{n}}\int_{0}^{1}p^{n-1-\sigma k_{n}}f_{\sigma}((1-p)t)\ddr p\ddr t.
\end{equation}
According to  \eqref{eq:gibbs_pred} and \eqref{eq_weight}, a Gibbs-type prior is parameterized by $(\sigma,h,\nu_{0})$; we denote by $Q_{h}$ this Gibbs-type random probability measure. The expression \eqref{eq:pred_pd} for the two-parameter PD prior is recovered from \eqref{eq_weight} by setting $h(t)=p(t;\sigma,\theta):=\sigma\Gamma(\theta)t^{-\theta}/\Gamma(\theta/\sigma)$, for any $\sigma\in(0,1)$ and $\theta>-\sigma$. The expression \eqref{eq:pred_pd} for the normalized GG prior is recovered from \eqref{eq_weight} by setting $h(t)=g(t;\sigma,\tau):=\exp\{\tau^{\sigma}-\tau t\}$, for any $\tau>0$. See Section 5.4 in \citet{Pit(03)} for details.

Besides providing a parameterization for Gibbs-type priors, the representation \eqref{eq_weight} leads to a simple numerical evaluation of $V_{h,(n,k_{n})}$. Specifically, let $B_{a,b}$ be a Beta random variable with parameter $(a,b)$ and, for any $\sigma\in(0,1)$ and $c>-1$, let $S_{\sigma,c}$ be a positive random variable with density function $f_{S_{\sigma,c}}(x)=\Gamma(c\sigma+1)x^{-c\sigma}f_{\sigma}(x)/\Gamma(c+1)$. $S_{\sigma,c}$ is  typically referred to as the polynomially tilted $\sigma$-stable random variable. Simple algebraic manipulations of \eqref{eq_weight} lead to
\begin{equation}\label{eq_weight_1}
V_{h,(n,k_{n})}=\frac{\sigma^{k_{n}-1}\Gamma(k_{n})}{\Gamma(n)}\E\left[h\left(\frac{S_{\sigma,k_{n}}}{B_{\sigma k_{n},n-\sigma k_{n}}}\right)\right],
\end{equation}
with $B_{\sigma k_{n},n-\sigma k_{n}}$ independent of $S_{\sigma,k_{n}}$. According to \eqref{eq_weight_1} a Monte Carlo evaluation of $V_{h,(n,k_{n})}$ can be performed by sampling from $B_{\sigma k_{n},n-\sigma k_{n}}$ and $S_{\sigma,k_{n}}$. In this respect, an efficient rejection sampling for $S_{\sigma,c}$ has been proposed by \citet{Dev(09)}. The next theorem, combined with \eqref{eq_weight_1}, provides a practical tool for obtaining an approximate evaluation of the credible intervals for $\hat{\mathcal{D}}_{n}(l)$.\smallskip\\

\textsc{Theorem 1.} Let $\boldsymbol{X}_{n}$ be a sample generated from $Q_{h}$ according to \eqref{eq:exchangeable_model} and featuring $K_{n}=k_{n}$ species, labelled by $X_{1}^{\ast},\ldots,X_{K_{n}}^{\ast}$, with corresponding frequencies $(N_{1,n},\ldots,N_{K_{n},n})=(n_{1,n},\ldots,n_{k_{n},n})$. For any set $A$ in the Borel sigma-algebra of $\mathbb{X}$, let $\mu_{n,k_{n}}(A)=\sum_{1\leq i\leq k_{n}}(n_{i,n}-\sigma)\delta_{X_{i}^{\ast}}(A)$. Then, for any $r\geq1$, the $r$th moment $\E[(Q_{h}(A))^{r}\,|\,\boldsymbol{X}_{n}]$ coincides with
\begin{align}\label{eq:rmom}
\sum_{i=0}^{r}\frac{V_{h,(n+r,k_{n}+i)}}{V_{h,(n,k_{n})}}(\nu_{0}(A))^{i} \sum_{0\leq j_{1}\leq\cdots\leq j_{i}\leq i}\prod_{q=0}^{r-i-1}(\mu_{n,k_{n}}(A)+j_{q}(1-\sigma)+q).
\end{align}
\smallskip

Let $\boldsymbol{M}_{n}:=(M_{1,n},\ldots,M_{n,n})=(m_{1,n},\ldots,m_{n,n})$ be the frequency counts from a sample $\boldsymbol{X}_{n}$ from $Q_{h}$. In order to obtain credible intervals for $\hat{\mathcal{D}}_{n}(l)$ we take two specifications of the Borel set $A$: $A_{0}=\mathbb{X}\setminus\{X_{1}^{\ast},\ldots,X_{K_{n}}^{\ast}\}$ and $A_{l}=\{X_{i}^{\ast}:N_{i,n}=l\}$, for any $l=1,\ldots,n$. With them, \eqref{eq:rmom} reduces to
\begin{gather}\label{eq:rmom_neq}
\mathcal{E}_{n,r}(0)=\E[(Q_{h}(A_{0}))^{r}\,|\,\boldsymbol{X}_{n}]=\sum_{i=0}^{r}{r\choose i}(-1)^{i}\frac{V_{h,(n+i,k_{n})}}{V_{h,(n,k_{n})}}(n-\sigma k_{n})_{i},\\
\label{eq:rmom_l}
\mathcal{E}_{n,r}(l)=\E[(Q_{h}(A_{l}))^{r}\,|\,\boldsymbol{X}_{n}]=\frac{V_{h,(n+r,k_{n})}}{V_{h,(n,k_{n})}}((l-\sigma)m_{l,n})_{r},
\end{gather}
respectively. Equations \eqref{eq:rmom_neq} and \eqref{eq:rmom_l} take on the interpretation of the $r$-th moments of the posterior distribution of $D_{n}(0)$ and $D_{n}(l)$ under the specification of a Gibbs-type prior. In particular for $r=1$, by using the recursion $V_{h,(n,k_{n})}=(n-\sigma k_{n})V_{h,(n+1,k_{n})}+V_{h,(n+1,k_{n}+1)}$, \eqref{eq:rmom_neq} and \eqref{eq:rmom_l} reduce to the Bayesian nonparametric estimators of $D_{n}(l)$ displayed resp. in \eqref{eq:disc_new} and \eqref{eq:disc_old}.

The distribution of $Q_{h}(A_{l})\,|\,\boldsymbol{X}_{n}$ is on $[0,1]$ and, therefore, it is characterized by $(\mathcal{E}_{n,r}(l))_{r\geq1}$. The approximation of a distribution given its moments is a longstanding problem which has been tackled by such approaches as expansions in polynomial bases, maximum entropy methods, and mixtures of distributions. For instance, the polynomial approach consists in approximating the density function of $Q_{h}(A_{l})\,|\,\boldsymbol{X}_{n}$ with a linear combination of orthogonal polynomials, where the coefficients of the combination are determined by equating $\mathcal{E}_{n,r}(l)$ with the moments of the approximating density. The higher the degree of the polynomials, or equivalently the number of moments used, the more accurate the approximation. As a rule of thumb, ten moments turn out to be enough in most cases. See \citet{Pro(05)} for details. The approximating density function of $Q_{h}(A_{l})\,|\,\boldsymbol{X}_{n}$ can then be used to obtain an approximate evaluation of the credible intervals for $\hat{\mathcal{D}}_{n}(l)$. This is typically done by generating random variates, via rejection sampling, from the approximating distribution of $Q_{h}(A_{l})\,|\,\boldsymbol{X}_{n}$. See \citet{arbel2014full}  for details.

Under the specification of the two-parameter PD prior and the normalized GG prior, \eqref{eq:rmom_neq} and \eqref{eq:rmom_l} lead to explicit  and simple characterizations for the distributions of $Q_{p}(A_{l})\,|\,\boldsymbol{X}_{n}$ and $Q_{g}(A_{l})\,|\,\boldsymbol{X}_{n}$, respectively. Let $G_{a,1}$ be a Gamma random variable with parameter $(a,1)$ and, for any $\sigma\in(0,1)$ and $b>0$, let $R_{\sigma,b}$ be a random variable with density function $f_{R_{\sigma,b}}(x)=\exp\{b^{\sigma}-bx\}f_{\sigma}(x)$. $R_{\sigma,b}$ is typically referred to as the exponentially tilted $\sigma$-stable random variable. Finally, define $W_{a,b}=bR_{\sigma,b}/(bR_{\sigma,b}+G_{a,1})$, 
where $G_{a,1}$ is independent of $R_{\sigma,b}$. The random variable $W_{a,b}$ is nonnegative and with values on the set $[0,1]$. 
\medskip

\textsc{Proposition 1.} 
Let $\boldsymbol{X}_{n}$ be a sample generated from $Q_{p}$ according to \eqref{eq:exchangeable_model} and featuring $K_{n}=k_{n}$ species with $\boldsymbol{M}_{n}=(m_{1,n},\ldots,m_{n,n})$. Let $Z_{p}$ be a nonnegative random variable with density function of the form
\begin{equation*}
f_{Z_{p}}(x)=\frac{\sigma}{\Gamma(\theta/\sigma+k_{n})}x^{\theta+\sigma k_{n}-1}\text{e}^{-x^{\sigma}}\mathds{1}_{(0,+\infty)}(x).
\end{equation*}
Then, $Q_{p}(A_{0})\,|\,\boldsymbol{X}_{n}\stackrel{\text{d}}{=}W_{n-\sigma k_{n},Z_{p}}\stackrel{\text{d}}{=}B_{\theta+\sigma k_{n},n-\sigma k_{n}}$
and 
$Q_{p}(A_{l})\,|\,\boldsymbol{X}_{n} \stackrel{\text{d}}{=}$\\ $B_{(l-\sigma)m_{l,n},n-\sigma k_{n}-(l-\sigma)m_{l,n}}(1-W_{n-\sigma k_{n},Z_{p}}) \stackrel{\text{d}}{=} B_{(l-\sigma)m_{l,n},\theta+n-(l-\sigma)m_{l,n}}$.
\medskip

\textsc{Proposition 2.} Let $\boldsymbol{X}_{n}$ be a sample generated from $Q_{g}$ according to \eqref{eq:exchangeable_model} and featuring $K_{n}=k_{n}$ species with $\boldsymbol{M}_{n}=(m_{1,n},\ldots,m_{n,n})$. Let  $Z_{g}$ be a nonnegative random variable with density function of the form
\begin{equation}\label{eq:latent_gg}
f_{Z_{g}}(x)=\frac{\sigma x^{\sigma k_{n}-n}(x-\tau)^{n-1}\exp\{-x^{\sigma}\}\mathds{1}_{(\tau,+\infty)}(x)}{\sum_{0\leq i\leq n-1}{n-1\choose i}(-\tau)^{i}\Gamma(k_{n}-i/\sigma;\tau^{\sigma})}.
\end{equation}
Then, 
$Q_{g}(A_{0})\,|\,\boldsymbol{X}_{n}\stackrel{\text{d}}{=}W_{n-\sigma k_{n},Z_{g}}$
 and 
$Q_{g}(A_{l})\,|\,\boldsymbol{X}_{n}\stackrel{\text{d}}{=}B_{(l-\sigma)m_{l,n},n-\sigma k_{n}-(l-\sigma)m_{l,n}}(1-W_{n-\sigma k_{n},Z_{g}})$.
\smallskip

According to Propositions 1 and 2, the random variables $Q_{p}(A_{0})\,|\,\boldsymbol{X}_{n}$ and $Q_{g}(A_{0})\,|\,\boldsymbol{X}_{n}$ have a common structure driven by the $W$ random variable. 
Moreover, for any $l=1,\ldots,n$, $Q_{p}(A_{l})\,|\,\boldsymbol{X}_{n}$ and $Q_{g}(A_{l})\,|\,\boldsymbol{X}_{n}$ are obtained by taking the same random proportion $B_{(l-\sigma)m_{l,n},n-\sigma k_{n}-(l-\sigma)m_{l,n}}$ of $(1-W_{n-\sigma k_{n},Z_{p}})$ and $(1-W_{n-\sigma k_{n},Z_{g}})$, respectively. Under the specification of the two-parameter PD prior and the normalized GG prior, Propositions 1 and 2 provide practical tools for  deriving credible intervals for the Bayesian nonparametric estimator $\hat{\mathcal{D}}_{n}(l)$, for any $l=0,1,\ldots,n$. This is typically done by performing a numerical evaluation of appropriate quantiles of the distribution of $Q_{p}(A_{l})\,|\,\boldsymbol{X}_{n}$ and $Q_{g}(A_{l})\,|\,\boldsymbol{X}_{n}$. In the special case of the Beta distribution, quantiles can be also determined explicitly as solutions of a certain class of non-linear ordinary differential equations. See \citet{Ste(08)} and references therein for a detailed account on this approach.

To obtain credible intervals for $\hat{\mathcal{D}}_{n}(l)$, we generate random variates from $Q_{p}(A_{l})\,|\,\boldsymbol{X}_{n}$ and $Q_{g}(A_{l})\,|\,\boldsymbol{X}_{n}$. With the two-parameter PD prior, sampling from $Q_{p}(A_{l})\,|\,\boldsymbol{X}_{n}$ for any $l=0,1,\ldots,n$ is straightforward, requiring generation of random variates from a Beta distribution. With the normalized GG prior, sampling from $Q_{p}(A_{l})\,|\,\boldsymbol{X}_{n}$ for any $l=0,1,\ldots,n$ is also straightforward. 
As the density function of  the transformed random variable $Z^{\sigma}_{g}$ is log-concave, one can sample from $Z^{\sigma}_{g}$ by means of the adaptive rejection sampling of \citet{Gil(92)}. Given $Z_{g}$, the problem of sampling from $W_{n-\sigma k_{n},Z_{g}}$ boils down to the problem of generating random variates from the distribution of the exponentially tilted $\sigma$-stable random variable $R_{\sigma,Z_{g}}$. This can be done by resorting to the efficient rejection sampling proposed by \citet{Dev(09)}.


\setcounter{equation}{0} 
\section{Large sample asymptotics for $\hat{\mathcal{D}}_{n}(l)$}\label{s:gt}

We investigate the large $n$ asymptotic behavior of the estimator $\hat{\mathcal{D}}_{n}(l)$, with a view towards its asymptotic relationships with smoothed Good--Turing estimators. Under a Gibbs-type prior, the most notable difference between the Good--Turing estimator $\check{\mathcal{D}}_{n}(l)$ and $\hat{\mathcal{D}}_{n}(l)$ can be traced to the different use of the information contained in the sample $\boldsymbol{X}_{n}$. Thus $\check{\mathcal{D}}_{n}(0)$ is a function of $m_{1,n}$ while $\hat{\mathcal{D}}_{n}(0)$ is a function of $k_{n}$, and $\check{\mathcal{D}}_{n}(l)$ is a function of $m_{l+1,n}$ while $\hat{\mathcal{D}}_{n}(l)$ is a function of $m_{l,n}$, for any $l=1,\ldots,n$. Let $a_{n}\simeq b_{n}$ mean that $\lim_{n\rightarrow+\infty}a_{n}/b_{n}=1$. 
We show that, as $n$ tends to infinity, $\hat{\mathcal{D}}_{n}(l)\simeq\check{\mathcal{D}}_{n}(l;\mathscr{S}_{\text{PD}})$, where $\mathscr{S}_{\text{PD}}$ is the smoothing rule displayed in \eqref{eq:smoother}. Such a result thus generalizes Theorem 1 in \citet{Fav(15)} to the entire class of Gibbs-type priors. The asymptotic results of this section hold almost surely, but the probabilistic formalization of this idea is postponed to the proofs in the Appendix.

\textsc{Theorem 2.} For almost every sample $\boldsymbol{X}_{n}$ generated from $Q_{h}$ according to \eqref{eq:exchangeable_model} and featuring $K_{n}=k_{n}$ species with $\boldsymbol{M}_{n}=(m_{1,n},\ldots,m_{n,n})$, we have
\begin{gather}\label{eq:asymp_0}
\hat{\mathcal{D}}_{n}(0)=\frac{\sigma k_{n}}{n}+o\left(\frac{k_{n}}{n}\right),\\
\label{eq:asymp_l}
\hat{\mathcal{D}}_{n}(l)=(l-\sigma)\frac{m_{l,n}}{n}+o\left(\frac{m_{l,n}}{n}\right).
\end{gather}
\medskip

By a direct application of Proposition 13 in \citet{Pit(03)} and Corollary 21 in \citet{Gne(07)} we can write that, for almost every sample $\boldsymbol{X}_{n}$ from $Q_{p}$, featuring $K_{n}=k_{n}$ species with $\boldsymbol{M}_{n}=(m_{1,n},\ldots,m_{n,n})$,
\begin{equation}\label{eq:prior_as_eq}
m_{l,n}\simeq\frac{\sigma(1-\sigma)_{l-1}}{l!}k_{n},
\end{equation}
as $n\rightarrow+\infty$. 
By suitably combining \eqref{eq:asymp_0} and \eqref{eq:asymp_l} with \eqref{eq:prior_as_eq}, we obtain
\begin{equation}\label{eq:turing_bayes_pd}
\hat{\mathcal{D}}_{n}(l)\simeq(l+1)\frac{m_{l+1,n}}{n}\simeq(l+1)\frac{\frac{\sigma(1-\sigma)_{l}}{(l+1)!}k_{n}}{n},
\end{equation}
for any $l=0,1,\ldots,n$. See the Appendix for details on \eqref{eq:turing_bayes_pd}. The first equivalence in \eqref{eq:turing_bayes_pd} shows that, as $n$ tends to infinity, $\hat{\mathcal{D}}_{n}(l)$ is asymptotically equal to the Good--Turing estimator $\check{\mathcal{D}}_{n}(l)$, whereas the second equivalence shows that, as $n$ tends to infinity, $\mathscr{S}_{\text{PD}}$ is a smoothing rule for the frequency counts $m_{l,n}$ in $\check{\mathcal{D}}_{n}(l)$. We refer to Section 2 in \citet{Fav(15)} for a relationship between the smoothing rule $\mathscr{S}_{\text{PD}}$ and the Poisson smoothing in \citet{Goo(53)}.

A peculiar feature of $\mathscr{S}_{\text{PD}}$ is that it does not depend on the function $h$ characterizing the Gibbs-type prior. 
Thus, for instance, $\mathscr{S}_{\text{PD}}$ is a smoothing rule for both the two-parameter PD prior and the normalized GG prior. This invariance property of $\mathscr{S}_{\text{PD}}$ is clearly determined by the fact that the asymptotic equivalences in  \eqref{eq:turing_bayes_pd} arise by combining \eqref{eq:prior_as_eq}, which does not depend on $h$, with \eqref{eq:asymp_0} and \eqref{eq:asymp_l}, which also do not depend of $h$. 
It is worth noticing that, unlike the smoothing rule $\mathscr{S}_{\text{PD}}$, the corresponding smoothed estimator $\check{\mathcal{D}}(l;\mathscr{S}_{\text{PD}})$ does depend on $h$ through $k_n$. Indeed, according to model \eqref{eq:exchangeable_model}, $Q$ is the data generating process and therefore the choice of a specific Gibbs-type prior $\mathcal{Q}$ or, in other terms, the specification of $h$, affects the distribution of $K_n$.
Intuitively, smoothing rules depending on the function $h$, if any exists, necessarily require to combine refinements of the asymptotic expansions \eqref{eq:asymp_0} and \eqref{eq:asymp_l} with corresponding refinements of the asymptotic equivalence \eqref{eq:prior_as_eq}. Under the specification of the two-parameter PD prior and the normalized GG prior, the next propositions provide asymptotic refinements of Theorem 2.\\

\textsc{Proposition 3.} For almost every sample $\boldsymbol{X}_{n}$ generated from $Q_{p}$ according to \eqref{eq:exchangeable_model} and featuring $K_{n}=k_{n}$ species with $\boldsymbol{M}_{n}=(m_{1,n},\ldots,m_{n,n})$, we have
\begin{gather*}
\hat{\mathcal{D}}_{n}(0)=\frac{\sigma k_{n}}{n}+\frac{\theta}{n}+o\left(\frac{1}{n}\right),\quad
\hat{\mathcal{D}}_{n}(l)=(l-\sigma)\frac{m_{l,n}}{n}\left(1-\frac{\theta}{n}\right)+o\left(\frac{m_{l,n}}{n^{2}}\right).
\end{gather*}
\smallskip

\textsc{Proposition 4.} For almost every sample $\boldsymbol{X}_{n}$ generated from $Q_{g}$ according to \eqref{eq:exchangeable_model} and featuring $K_{n}=k_{n}$ species with $\boldsymbol{M}_{n}=(m_{1,n},\ldots,m_{n,n})$, we have 
\begin{equation*}
\hat{\mathcal{D}}_{n}(0)=\frac{\sigma k_{n}}{n}+\tau k_{n}^{-1/\sigma}+o\left(\frac{1}{n}\right),
\quad
\hat{\mathcal{D}}_{n}(l)=(l-\sigma)\frac{m_{l,n}}{n}\left(1-\tau k_{n}^{-1/\sigma}\right)+o\left(\frac{m_{l,n}}{n^{2}}\right).
\end{equation*}

\medskip

In Propositions 3 and 4, we introduce second order approximations of $\hat{\mathcal{D}}_{n}(0)$ and $\hat{\mathcal{D}}_{n}(l)$ by considering a two-term truncation of the corresponding asymptotic series expansions. Here it is sufficient to include the second term in order to introduce the dependency on $\theta>-\sigma$ and $\tau>0$, respectively, and then the approximations of $\hat{\mathcal{D}}_{n}(0)$ and $\hat{\mathcal{D}}_{n}(l)$ differ between the two-parameter PD  prior and the normalized GG prior.

The second order approximations in Propositions 3 and 4, in combination with corresponding second order refinements of \eqref{eq:prior_as_eq}, do not lead to a second order refinement of \eqref{eq:turing_bayes_pd}. 
A second order refinement of  \eqref{eq:prior_as_eq}, arising from \citet{Gne(07)}, can be expressed as 
\begin{equation}\label{eq:Gnedin_refinement}
M_{l,n} = \frac{\sigma(1-\sigma)_{l-1}}{l!}K_{n}+O\left(\frac{K_{n}}{n^{\sigma/2}}\right),
\end{equation}
but second order terms in Propositions 3 and 4 are absorbed by $O\left(K_{n}/n^{\sigma/2}\right)$ in \eqref{eq:Gnedin_refinement}. Furthermore, even if a finer version of \eqref{eq:Gnedin_refinement} was available, its combination with Propositions 3 and 4 would produce higher order terms preventing the resulting expression from being interpreted as a Good--Turing estimator and, therefore, any smoothing rule from being elicited. In other terms, under the two-parameter PD and the normalized GG priors, the relationship between $\hat{\mathcal{D}}_{n}(l)$ and $\check{\mathcal{D}}_{n}(l)$ only holds at the order of asymptotic equivalence. Theorem 2 and Proposition 4, as to the normalized GG prior, provide useful approximations that might dramatically fasten up the evaluation of $\hat{\mathcal{D}}_n(l)$, for $l=0,1,\ldots,n$, when $n$ is large, by avoiding the  Monte Carlo evaluation of the $V_{n,k_n}$'s appearing in \eqref{eq:disc_new} and \eqref{eq:disc_old}.


\setcounter{equation}{0} 
\section{Illustrations}\label{s:ill}

We illustrate our results with simulations and analysis of data. Data were generated from the Zeta distribution, whose power law behavior is common in a variety of applications. See \citet{Sam(01)} and references therein for applications of the Zeta distribution in empirical linguistics. One has $\P[Z=z]=z^{-s}/C(s)$, for $z=\{1,2,\ldots\}$ and $s>1$, where $C(s) =\sum_{i\geq1} i^{-s}$. We took $s=1.1$ (case $s=1.5$, typically leading to samples with a smaller number of distinct values, is presented in the Appendix).
We drew 500 samples of size $n=1,\,000$ from $Z$, ordered them according to the number of observed species $k_n$, and split them into 5 groups: for $i=1,2,\ldots,5$, the $i$-th group of samples was composed of 100 samples featuring a total number of observed species $k_n$  between the quantiles of order $(i-1)/5$ and $i/5$ of the empirical distribution of $k_n$. Then we chose at random one sample for each group and labeled it with the corresponding index $i$, leading to five samples (see Table~\ref{table_samples}).

We also considered ESTs data generated by sequencing two \emph{Naegleria gruberi} complementary DNA libraries; these were prepared from cells grown under different culture conditions, aerobic and anaerobic conditions. The rate of gene discovery depends on the degree of redundancy of the library from which such sequences are obtained. Correctly estimating the relative redundancy of such libraries, as well as other quantities such as the probability of sampling a new or a rarely observed gene, is of importance since it allows one to optimize the use of expensive experimental sampling techniques. The \emph{Naegleria gruberi} aerobic library consists of $n = 959$ ESTs with $k_n = 473$ distinct genes and $m_{l,959} = 346,57,19,12,9,5,4,2,4,5,4,1,1,1,1,1,1$, for $l= \{1,2,\ldots,12\}\cup\{16,17,18\}\cup\{27\}\cup\{55\}$. The \emph{Naegleria gruberi} anaerobic library consists of $n = 969$ ESTs with $k_n = 631$ distinct genes and $m_{l,969}=491,72,30,9,13,5,3,1,2,0,1,0,1$, for $l\in\{1,2,\ldots,13\}$ (see Table~\ref{table_samples}). We refer to \citet{Sus(04)} for a detailed account on the \emph{Naegleria gruberi} libraries.

We focused on the two-parameter PD prior and the normalized GG prior. 
We choose the values of $(\sigma,\theta)$ and $(\sigma,\tau)$ by an empirical Bayes approach, as those that  maximized the likelihood function with respect to the sample $\boldsymbol{X}_{n}$ featuring $K_{n}=k_{n}$ and $(N_{1,n},\ldots,N_{K_{n},n})=(n_{1,n},\ldots,n_{k_{n},n})$, 
\begin{gather}\label{eq:eb_pd}
(\hat\sigma,\hat\theta)=\operatorname*{arg\,max}_{(\sigma,\theta)}\left\{\frac{\prod_{i=0}^{{k_n}-1}(\theta+i\sigma)}{(\theta)_{n}}\prod_{i=1}^{{k_n}}(1-\sigma)_{(n_{i,n}-1)}\right\},\\
\label{eq:eb_gg}
(\hat\sigma,\hat \tau)=\operatorname*{arg\,max}_{(\sigma,\tau)}\left\{\frac{\edr^{\tau^\sigma} \sigma^{{k_n}-1}}{\Gamma(n)} \sum_{i=0}^{n-1}\binom{n-1}{i}(-\tau)^{i}\Gamma\left({k_n}-\frac{i}{\sigma};\tau^\sigma\right)\prod_{i=1}^{{k_n}}(1-\sigma)_{(n_{i,n}-1)}\right\}.
\end{gather} 
As first observed by \citet{favaro2009bayesian}, under the specification of the two-parameter PD prior and for a relatively large observed sample, there is a high concentration of the posterior distribution of the parameter $(\sigma,\theta)$ around $(\hat{\sigma},\hat{\theta})$. It can be checked that, under the specification of a normalized GG prior, a similar behaviour characterizes the posterior distribution of $(\sigma,\tau)$. 

Table \ref{table_samples} reports the sample size $n$, the number of species $k_{n}$, and the values of $(\hat{\sigma},\hat{\theta})$ and $(\hat{\sigma},\hat{\tau})$ obtained by the maximizations \eqref{eq:eb_pd} and \eqref{eq:eb_gg}, respectively. Here the value of $\hat{\sigma}$ obtained under the two-parameter PD prior coincides, up to a negligible error, with the value of $\hat{\sigma}$ obtained under the normalized GG prior. In general, we expect the same behaviour for any Gibbs-type prior in light of the likelihood function of a sample $\boldsymbol{X}_{n}$ from a Gibbs-type random probability measure $Q_{h}$, 
\begin{equation}\label{eq:like_gibbs}
\frac{\sigma^{k_{n}}\prod_{i=1}^{k_{n}}(1-\sigma)_{(n_{i}-1)}}{\Gamma(n-\sigma k_{n})}\int_{0}^{+\infty}h(t)t^{-\sigma k_{n}}\int_{0}^{1}p^{n-1-\sigma k_{n}}f_{\sigma}((1-p)t)\ddr p\ddr t.
\end{equation}
Apart from $\sigma$, any other parameter is introduced in \eqref{eq:like_gibbs} via the function $h$, which does not depend on the sample size $n$ and the number of species $k_{n}$. Then, for large $n$ and $k_{n}$ the maximization of \eqref{eq:like_gibbs} with respect to $\sigma$ should lead to a value $\hat{\sigma}$ very close to the value that would be obtained by maximizing \eqref{eq:like_gibbs} with $h(t)=1$. 
\begin{table}
\caption{Simulated data and \emph{Naegleria gruberi} libraries. For each sample we report the sample size $n$,  number of species ${k_n}$ and maximum likelihood values $(\hat \sigma, \hat \theta)$ and $(\hat \sigma, \hat \tau)$.}
      \label{table_samples}
   \begin{small}
      \begin{center}
      \begin{tabular}{|cccc|cc|cc|}
      \hline
         \multicolumn{4}{|c|}{} &  \multicolumn{2}{c|}{PD} & \multicolumn{2}{c|}{GG} \\
      \hline
      & sample & $n$ & ${k_n}$ & $\hat\sigma$ & $\hat\theta$ & $\hat \sigma$ & $\hat\tau$ \\ 
    \hline
      \multirow{5}{*}{Simulated data}& $ 1$ & 1,\,000 &  642  &     0.914  &  2.086 &     0.913  &2.517 \\
      & $ 2$ & 1,\,000  &   650   &     0.905 &   3.812 &     0.905  & 4.924\\
      & $ 3$ &  1,\,000  & 656 &     0.910  &  3.236 &    0.910  & 4.060\\
      & $ 4$ &  1,\,000  &   663 &     0.916 &   2.597 &    0.916  & 3.156\\
      & $ 5$ &  1,\,000  &   688  &      0.920  &  3.438 &     0.920  & 4.225\\
    \hline
          \multirow{2}{*}{Naegleria} & Aerobic & 959 & 473 & 0.669  & 46.241 & 0.684 & 334.334 \\    
      & Anaerobic &  969 & 631 & 0.656 & 155.408 & 0.656 & 4151.075\\
    \hline 
      \end{tabular}
      \end{center}
 \end{small}
\end{table}
\subsection{Credible intervals}\label{s:ill1}
We applied Propositions 1 and 2 in order to provide  credible intervals for  the Bayesian nonparametric estimator  $\hat{\mathcal{D}}_{n}(l)$. For the two-parameter PD prior, for $l=0$ we generated 5,\,000 draws from the beta  $B_{\hat\theta+\hat\sigma k_{n},n-\hat\sigma k_{n}}$ while, for $l\geq1$ we sampled 5,\,000 draws from the distribution of a beta random variable $B_{(l-\hat\sigma)m_{l,n},\hat\theta+n-(l-\hat\sigma)m_{l,n}}$. In both cases, we computed the quantiles of order $\{0.025, 0.975\}$ of the empirical distribution and obtained $95\%$ posterior credible intervals for $\hat{\mathcal{D}}_{n}(l)$. The procedure for the normalized GG case was only slightly more elaborate. By exploiting the adaptive rejection algorithm of \citet{Gil(92)}, we sampled 5,\,000 draws from $Z_{g}$ with density function \eqref{eq:latent_gg}. In turn, we sampled 5,\,000 draws from $W_{n-\hat\sigma k_n, Z_g}$. We then used the quantiles of order $\{0.025, 0.975\}$ of the empirical distribution of $W_{n-\hat\sigma k_n, Z_g}$ to obtain $95\%$ posterior credible intervals for $\hat{\mathcal{D}}_{n}(0)$. Similarly, if $l\geq 1$, we sampled $5,\,000$ draws from the beta $B_{(l-\hat\sigma)m_{l,n},n-\hat\sigma k_{n}-(l-\hat\sigma)m_{l,n}}$ and used the quantiles of the empirical distribution of $B_{(l-\hat\sigma)m_{l,n},n-\hat\sigma k_{n}-(l-\hat\sigma)m_{l,n}}(1-W_{n-\hat\sigma k_n, Z_g})$ as extremes of the posterior credible interval for $\hat{\mathcal{D}}_{n}(l)$.
\begin{table}
      \caption{Simulated data (top panel) and \emph{Naegleria gruberi} aerobic and anaerobic libraries (bottom panel). We report the true value of the probability $D_{n}(l)$ (available for simulated data only) and the Bayesian nonparametric estimates of $D_{n}(l)$ with 95\% credible intervals for $l=0,1,5,10$.}
      \label{table_samples_2}
      \begin{small}
      \begin{center}
      \begin{tabular}{|ccc|cc|cc|cc|}
      \hline 
         \multicolumn{3}{|c|}{} & \multicolumn{2}{c|}{Good--Turing} &  \multicolumn{2}{c|}{PD} & \multicolumn{2}{c|}{GG}   \\
      \hline
           $l$  & sample & $D_{n}(l)$ & $\check{\mathcal{D}}_n(l)$ & 95\%-c.i. & $\hat{\mathcal{D}}_n(l)$ & 95\%-c.i. & $\hat{\mathcal{D}}_n(l)$ & 95\%-c.i.  \\
    \hline
      \multirow{5}{*}{$0$}
      & $1$ & 0.599 & 0.588 & (0.440, 0.736) & 0.587 & (0.557, 0.618)  & 0.588 & (0.558, 0.620) \\
      & $2$ & 0.592 & 0.590 & (0.454, 0.726) & 0.590 & (0.559, 0.621)  & 0.591 & (0.562, 0.620) \\
      & $3$ & 0.600 & 0.599 & (0.462, 0.736) & 0.598 & (0.568, 0.628)  & 0.599 & (0.567, 0.630) \\
      & $4$ & 0.605 & 0.609 & (0.473, 0.745) & 0.609 & (0.579, 0.638)  & 0.608 & (0.577, 0.638) \\
      & $5$ & 0.599 & 0.634 & (0.499, 0.769) & 0.634 & (0.603, 0.664)  & 0.635 & (0.604, 0.663) \\
     \hline      
          \multirow{5}{*}{$1$} 
      & $1$ & 0.050 & 0.044 & (0.037, 0.051)  & 0.051 & (0.038, 0.065)  & 0.051 & (0.038, 0.065) \\
      & $2$ & 0.052 & 0.054 & (0.046, 0.062) & 0.056  & (0.043, 0.071) & 0.055 & (0.042, 0.070) \\
      & $3$ & 0.051 & 0.046 & (0.039, 0.053)  & 0.054 & (0.040, 0.068)  & 0.053 & (0.040, 0.068) \\
      & $4$ & 0.055 & 0.046 & (0.039, 0.053) & 0.051 & (0.038, 0.065)  & 0.051 & (0.038, 0.065) \\
      & $5$ & 0.061 & 0.052 & (0.045, 0.059)  & 0.051 & (0.038, 0.065)  & 0.050 & (0.038, 0.064) \\
    \hline 
              \multirow{5}{*}{$5$} 
      & $1$ & 0.015 &   0.030  & (0.022, 0.038) & 0.016 & (0.009, 0.025)  & 0.016 & (0.009, 0.025) \\
      & $2$ & 0.022 & 0 & (0, 0)  & 0.016 & (0.009, 0.025)  & 0.016 & (0.009, 0.025) \\
      & $3$ & 0.019 & 0.012 & (0.008, 0.016) & 0.020 & (0.013, 0.030)  & 0.021 & (0.012, 0.030) \\
      & $4$ & 0.015 & 0.006 & (0.003, 0.009) & 0.020 & (0.013, 0.030)  & 0.021 & (0.013, 0.031) \\
      & $5$ & 0.007 & 0.012 & (0.007, 0.017) & 0.008 & (0.004, 0.015)  & 0.008 & (0.003, 0.015) \\
    \hline 
              \multirow{5}{*}{$10$} 
      & $1$ & 0 & 0.011 & n.a. &  0 & (0, 0)  & 0 & (0, 0) \\
      & $2$ & 0.007 & 0 & (0, 0) & 0.009 & (0.004, 0.016)  & 0.009 & (0.004, 0.016) \\
      & $3$ & 0.011 & 0 & (0, 0) &  0.009  & (0.004, 0.016) & 0.009 & (0.004, 0.016) \\
      & $4$ & 0.011 & 0 & (0, 0) & 0.009 & (0.004, 0.016)  & 0.009 & (0.004, 0.016) \\
      & $5$ & 0 & 0.011 & n.a. &  0 & (0, 0)  & 0 & (0, 0) \\
    
          \hline
    \hline
      \multirow{2}{*}{$0$}
      & Aerobic & n.a.  & 0.361 & (0.293, 0.429) & 0.361 & (0.331, 0.391)  & 0.361  & (0.332, 0.389) \\
      & Anaerobic & n.a. & 0.507 & (0.451, 0.562) & 0.509 & (0.478, 0.537)  & 0.507 & (0.480, 0.532) \\
          \hline      
          \multirow{2}{*}{$1$} 
      & Aerobic & n.a. & 0.119 & (0.107, 0.131) &  0.114 & (0.095, 0.134)  & 0.110 & (0.092, 0.131) \\
      & Anaerobic & n.a. & 0.149 & (0.135, 0.162) &  0.148 & (0.129, 0.169)  & 0.150 & (0.131, 0.172) \\
       \hline 
              \multirow{2}{*}{$5$} 
      & Aerobic & n.a.  & 0.031 & (0.024, 0.038) &  0.039 & (0.028, 0.052)  & 0.039 & (0.028, 0.053) \\
      & Anaerobic & n.a. & 0.031 & (0.024, 0.038) &  0.050 & (0.038, 0.064)  & 0.050 & (0.038, 0.064) \\
     \hline 
              \multirow{2}{*}{$10$} 
      & Aerobic & n.a.  & 0.046 & (0.037, 0.055) &  0.046 & (0.034, 0.060)  & 0.047 & (0.034, 0.061) \\
      & Anaerobic & n.a.  & 0.011 & n.a. &  0 & (0, 0)  & 0 & (0, 0) \\
      \hline
      \end{tabular}
      \end{center}
  \end{small}
\end{table}
Under the two-parameter PD prior and the normalized GG prior, and with respect to these data, the top panel of Table \ref{table_samples_2} 
shows the estimated $l$-discoveries, for $l=0,1,5,10$, and the corresponding $95\%$ posterior credible intervals. It is apparent that the two-parameter PD prior and the normalized GG prior lead to the same inferences for the $l$-discovery. Such a behaviour is mainly determined by the fact that the two-parameter PD prior, for any $\sigma\in(0,1)$ and $\theta>0$, can be viewed as a mixture of normalized GG priors. Specifically, let $\mathscr{Q}_{p}(\sigma,\theta)$ and $\mathscr{Q}_{g}(\sigma,b)$ be the distributions of the corresponding random probability measures, and let $G_{\theta/\sigma,1}$ be a Gamma random variable with parameter $(\theta/\sigma,1)$. Then, according to Proposition 21 in \citet{Pit(97)}, $\mathscr{Q}_{p}(\sigma,\theta)=\mathscr{Q}_{g}(\sigma,G^{1/\sigma}_{\theta/\sigma,1})$, and specifying a two-parameter PD prior is equivalent to specifying a normalized GG prior with an Gamma hyper prior over the parameter $\tau^{1/\sigma}$. 
Table~\ref{table_samples_2} allows us to compare the performance of the Bayesian nonparametric estimator $\hat{\mathcal{D}}_{n}(l)$ and the Good--Turing estimator $\check{\mathcal{D}}_{n}(l)$.
As expected, Good--Turing estimates are not reliable as soon as $l$ is not very small compared to $n$. See, e.g., the cases $l=5$ and $l=10$. Of course these estimates may be improved by introducing a suitable smoothing rule for the frequency counts $m_{l,n}$'s. We are not aware of a non-asymptotic approach for devising confidence intervals for $\check{\mathcal{D}}_{n}(l)$, and found that different procedures are used according to the choice of $l=0$ and $l\geq1$. We relied on \citet{Mao(04)} for $l=0$ and on \citet{Chu(91)} for $l\geq1$. See also \citet{Baa(01)} for details. We observe that the confidence intervals for $\check{\mathcal{D}}_{n}(l)$ are wider than the corresponding credible intervals for $\hat{\mathcal{D}}_{n}(l)$ when $l=0$, and narrower if $l\geq 1$. Differently from the credible intervals for $\hat{\mathcal{D}}_{n}(l)$, the confidence intervals for $\check{\mathcal{D}}_{n}(l)$ are symmetric about $\check{\mathcal{D}}_{n}(l)$; such a behaviour is determined by the Gaussian approximation used to derive confidence intervals.

\subsection{Large sample approximations}\label{s:ill2}

We analyzed the accuracy of the large $n$ approximations of $\hat{\mathcal{D}}_{n}(l)$ introduced in Theorem 2, Propositions 3 and 4. We first compared the precision of exact and approximated estimators, while a second analysis compared the behavior of first and second order approximations for varying sample sizes. For the simulated data, the specification of the two-parameter PD prior and the normalized GG prior, and for $l=0,1,5,10$, we compared the true discovery probabilities $D_{n}(l)$ with the Bayesian nonparametric estimates of $D_{n}(l)$ and with their corresponding first and second order approximations. 
From Table \ref{table_samples}, the empirical Bayes estimates for $\sigma$ can be slightly different under the two-parameter PD and the normalized GG priors. We considered only the first order approximation of $\hat{\mathcal{D}}_{n}(l)$ with the parameter $\sigma=\hat\sigma$ set as indicated in \eqref{eq:eb_pd}.

\begin{table}[h!p!t!]
      \caption{Simulated data. We report the true value of the probability $D_{n}(l)$, the Good--Turing estimates of $D_{n}(l)$ and the exact and approximate Bayesian nonparametric estimates of  $D_{n}(l)$.}
      \label{table_approx}
      \begin{small}
  \begin{center}
      \begin{tabular}{|c|c|ccccc|}
      \hline
   $l$ & Sample & 1 & 2 & 3 & 4 & 5\\ 
      \hline
      \multirow{7}{*}{$0$} &$D_{n}(l)$ & 0.599 & 0.592 & 0.600 & 0.605 & 0.599\\
      & $\check{\mathcal{D}}_{n}(l)$ & 0.588 & 0.590 & 0.599 & 0.609 & 0.634\\
      & $\hat{\mathcal{D}}_{n}(l)$ under PD & 0.587 & 0.590 & 0.598 & 0.609 & 0.634 \\
      & $\hat{\mathcal{D}}_{n}(l)$ under GG & 0.588 & 0. 591 & 0.599 & 0.608 & 0.635\\
       & 1st ord. & 0.587  & 0.588 & 0.597 & 0.608 & 0.633 \\
      & 2nd ord. PD  & 0.589  & 0.592 & 0.600 & 0.610  & 0.6366 \\
      & 2nd ord. GG  & 0.589 & 0.592 & 0.600 & 0.610 & 0.636\\[3pt]
      \hline
      \multirow{7}{*}{$1$} &$D_{n}(l)$ & 0.050 & 0.052 & 0.051 & 0.055 & 0.061 \\
      & $\check{\mathcal{D}}_{n}(l)$ & 0.044 & 0.054 & 0.046 & 0.046 & 0.052\\
      & $\hat{\mathcal{D}}_{n}(l)$ under PD & 0.051 & 0.056 & 0.054 & 0.051 & 0.051\\
      & $\hat{\mathcal{D}}_{n}(l)$ under GG & 0.051 & 0.055 & 0.053 & 0.051 & 0.050 \\
       & 1st ord. &  0.051 & 0.056 & 0.054   &    0.051 &    0.051 \\
      & 2nd ord. PD  &  0.051 & 0.056 & 0.054 &   0.051 &    0.051 \\
      & 2nd ord. GG  & 0.051 & 0.056 &   0.054 &    0.051 &    0.0512\\[3pt]
            \hline
      \multirow{7}{*}{$5$} &$D_{n}(l)$ & 0.015 & 0.022 & 0.019 & 0.015 & 0.007 \\
      & $\check{\mathcal{D}}_{n}(l)$ & 0.030 & 0 & 0.012 & 0.006 & 0.012\\
      & $\hat{\mathcal{D}}_{n}(l)$ under PD & 0.016 & 0.016 & 0.020 & 0.020 & 0.008\\
      & $\hat{\mathcal{D}}_{n}(l)$ under GG & 0.016 & 0.016 & 0.021 & 0.021 & 0.008\\
       & 1st ord. &  0.016  &  0.016 & 0.020 & 0.020 &    0.008 \\
      & 2nd ord. PD  & 0.016 & 0.016 & 0.020 &     0.020 &    0.008 \\
      & 2nd ord. GG  &   0.016 &   0.016 & 0.020 &    0.020 &   0.008\\[3pt]
            \hline
      \multirow{7}{*}{$10$} &$D_{n}(l)$ & 0 & 0.007 & 0.011 & 0.011 & 0 \\
      & $\check{\mathcal{D}}_{n}(l)$ & 0.011 & 0 & 0 & 0 & 0.011\\
      & $\hat{\mathcal{D}}_{n}(l)$ under PD & 0 & 0.009 & 0.009 & 0.009 & 0\\
      & $\hat{\mathcal{D}}_{n}(l)$ under GG & 0 & 0.009 & 0.009 & 0.009 & 0\\
       & 1st ord. & 0 &    0.009 &    0.009 &     0.009 & 0 \\
      & 2nd ord. PD  & 0 & 0.009 & 0.009 &    0.009 &         0 \\
      & 2nd ord. GG  & 0 &    0.009 &    0.009 &    0.009 &  0\\[3pt]
         \hline
                  \hline
       \multicolumn{2}{|c|}{$10^4\times\text{SSE}(\check{\mathcal{D}}_{n})$}& 289.266 & 275.881 & 256.886 & 254.416 & 255.655\\
       \multicolumn{2}{|c|}{$10^4\times\text{SSE}(\hat{\mathcal{D}}_{n})\text{ under PD}$}& 3.534 & 2.057 & 1.137 & 4.883 & 15.437 \\
       \multicolumn{2}{|c|}{$10^4\times\text{SSE}(\hat{\mathcal{D}}_{n})\text{ under GG}$}& 3.399 & 2.080 & 1.149 & 4.852 & 15.045\\
       \multicolumn{2}{|c|}{$10^4\times\text{SSE}(\hat{\mathcal{D}}_{n})\text{ 1st ord.}$}& 3.780 & 2.142 & 1.180 & 4.776 & 14.456 \\
       \multicolumn{2}{|c|}{$10^4\times\text{SSE}(\hat{\mathcal{D}}_{n})\text{ 2st ord. PD}$}& 3.275 & 2.011 & 1.128 & 5.041 & 17.007\\
              \multicolumn{2}{|c|}{$10^4\times\text{SSE}(\hat{\mathcal{D}}_{n})\text{ 2st ord. GG}$}& 3.279 & 2.014 & 1.130 & 5.035 & 16.984 \\[3pt]
       \hline
      \end{tabular}
      \end{center}
      \end{small}
\end{table}

Results of this comparative study are reported in Table \ref{table_approx}. We also include, as an overall measure of the performance of the exact and approximate estimators, the sum of squared errors (SSE), defined, for a generic estimator $\hat{D}_{n}(l)$ of the $l$-discovery, as $\text{SSE}(\hat{D}_{n})=\sum_{0\leq l\leq n}(\hat{D}_{n}(l)-d_{n}(l))^2$, with $d_n(l)$ being the true value of $D_{n}(l)$. For all the considered samples, there are not substantial differences between the SSEs of the exact Bayesian nonparametric estimates and the SSEs of the first and second order approximate Bayesian nonparametric estimates. The first order approximation is already pretty accurate and, thus, the approximation error does not contribute significantly to increase the SSE. As expected, the order of magnitude of the SSE referring to the not-smoothed Good--Turing estimator is much larger than the one corresponding to the Bayesian nonparametric estimators.

We considered simulated data with sample sizes  $n=10^2,10^3, 10^4,10^5$. For every $n$, we drew ten samples from a Zeta distribution with parameter $s=1.1$. We focused on the two-parameter PD prior, and for each sample we determined $(\hat\sigma,\hat\theta)$ by means of the empirical Bayes procedure described in \eqref{eq:eb_pd}. We then evaluated, for every $l=0,1,\ldots,n+1$, the exact estimator $\hat{\mathcal{D}}_{n}(l)$ as well as its first and second order approximations. To compare the relative accuracy of the first and second order approximations $\hat{\mathcal{D}}^{(1)}_{n}(l)$ and $\hat{\mathcal{D}}^{(2)}_{n}(l)$ of the same estimator $\hat{\mathcal{D}}_{n}(l)$ we introduce the ratio $r_{1,2,n}$ of the sum of squared errors $\sum_{0\leq l \leq n}(\hat{\mathcal{D}}^{(i)}_{n}(l)-\hat{\mathcal{D}}_{n}(l))^2$ for $i=1$ over $i=2$. 
We computed the coefficient $r_{1,2,n}$ for all the samples and, for each $n$, the average ratio $\bar{r}_{1,2,n}$. We found the increasing values $\bar{r}_{1,2,n} = 0.163, 0.493, 1.082, 2.239$ for sizes $n=10^2,10^3, 10^4,10^5$ (see Figure S1 in the  Appendix). While for small $n$ a first order approximation turns out to be more accurate, for large values of $n$ ($n\geq 10^4$ in our illustration), as expected, the second order approximation is more precise. 




\appendix

\section{Appendix}

This appendix contains: i) the proofs of Theorem 1, Proposition 1, Proposition 2, Theorem 2, Proposition 3 and Proposition 4; ii) details on the derivation of the asymptotic equivalence between $\hat{\mathcal{D}}_{n}(l)$ and  $\check{\mathcal{D}}_{n}(l;\mathscr{S}_{\text{PD}})$; iii) additional application results.


\setcounter{section}{0}
\setcounter{equation}{0}
\def\theequation{A\arabic{section}.\arabic{equation}}
\def\thesection{A\arabic{section}}

Let $\boldsymbol{X}_{n}=(X_{1},\ldots,X_{n})$ be a sample from a Gibbs-type RPM $Q_{h}$. Recall that, due to the discreteness of $Q_{h}$, the sample $\boldsymbol{X}_{n}$ features $K_{n}=k_{n}$ species, labelled by $X_{1}^{\ast},\ldots,X_{K_{n}}^{\ast}$, with corresponding frequencies $(N_{1,n},\ldots,N_{K_{n},n})=(n_{1,n},\ldots,n_{k_{n},n})$. Furthermore, let $M_{l,n}=m_{l,n}$ be the number of species with frequency $l$, namely $M_{l,n}=\sum_{1\leq i\leq K_{n}}\mathds{1}_{\{N_{i,n}=l\}}$ such that $\sum_{1\leq i\leq n}M_{i,n}=K_{n}$ and  $\sum_{1\leq i\leq n}iM_{i,n}=n$. For any $\sigma\in(0,1)$ let $f_{\sigma}$ be the density function of a positive $\sigma$-stable random variable. According to  Proposition 13 in \citet{Pit(03)}, as $n\rightarrow+\infty$
\begin{equation}\label{eq:asymp_dist}
\frac{K_{n}}{n^{\sigma}}\stackrel{\text{a.s.}}{\longrightarrow}S_{\sigma,h}
\end{equation}
and
\begin{equation}\label{eq:asymp_dist_freq}
\frac{M_{l,n}}{n^{\sigma}}\stackrel{\text{a.s.}}{\longrightarrow}\frac{\sigma(1-\sigma)_{l-1}}{l!}S_{\sigma,h},
\end{equation}
where $S_{\sigma,h}$ is a random variable with density function $f_{S_{\sigma,h}}(s)=\sigma^{-1}s^{-1/\sigma-1}h(s^{-1/\sigma})f_{\sigma}(s^{-1/\sigma})$. Note that by the fluctuation limits displayed in \eqref{eq:asymp_dist} and \eqref{eq:asymp_dist_freq}, as $n$ tends to infinity the number of species with frequency $l$ in a sample of size $n$ from $Q_{h}$ becomes, almost surely, a proportion $\sigma(1-\sigma)_{l-1}/l!$ of the total number of species in the sample. All the random variables introduced in this Appendix  are meant to be assigned on a common probability space $(\Omega,\mathscr{F},\P)$.

\section{Proofs}
\setcounter{equation}{0}

\textsc{Proof of Theorem 1.} We proceed by induction. Note that the result holds for $r=1$, and obviously for any sample size $n\geq 1$. Let us assume that it holds for a given $r\geq 1$, and also for any sample size $n\geq 1$. Then, the $(r+1)$-th moment of $Q_h(A)\,|\,\boldsymbol{X}_{n}$ can be written as follows
\begin{align*}
&\E[Q_h^r(A)\,|\,\boldsymbol{X}_{n}]\\
&\quad=\int_{A}\cdots\int_{A}\P[X_{n+r+1}\in A\,|\,\boldsymbol{X}_{n},X_{n+1}=x_{n+1},\ldots,X_{n+r}=x_{n+r}]\\
&\quad\quad\times\P[X_{n+r}\in \ddr x_{n+r}\,|\,\boldsymbol{X}_{n},X_{n+1}=x_{n+1},\ldots,X_{n+r-1}=x_{n+r-1}]\\
&\quad\quad\quad\times\cdots\times\P[X_{n+2}\in \ddr x_{n+2}\,|\,\boldsymbol{X}_{n},X_{n+1}=x_{n+1}]\P[X_{n+1}\in \ddr x_{n+1}\,|\,\boldsymbol{X}_{n}]\\
&\quad=\int_{A}\E[Q_h^r(A)\,|\,\boldsymbol{X}_{n},X_{n+1}=x_{n+1}] \\
&\quad\quad\times\left(\frac{\Vnuku}{\Vnk}\nu_{0}(\ddr x_{n+1})+\frac{\Vnuk}{\Vnk}\sum_{i=1}^{{k_n}}(n_{i}-\sigma)\delta_{X_{i}^{\ast}}(\ddr x_{n+1})\right).
\end{align*} 
Further, by the assumption on the $r$-th moment and by dividing $A$ into $(A\setminus\boldsymbol{X}_{n})\cup(A\cap\boldsymbol{X}_{n})$, one obtains
\begin{align*}
&\E[Q_h^{r+1}(A)\,|\,\boldsymbol{X}_{n}]\\
&\quad= \sum_{i=0}^{r}\frac{V_{n+r+1,{k_n}+r+1-i}}{\Vnk}[\nu_{0}(A)]^{r+1-i}R_{r,i}(\mu_{n,k_n}(A)+1-\sigma)\\
&\quad\quad + \sum_{i=1}^{r+1}\frac{V_{n+r+1,{k_n}+r+1-i}}{\Vnk}[\nu_{0}(A)]^{r+1-i}\mu_{n,k_n}(A)R_{r,i-1}(\mu_{n,k_n}(A)+1),
\end{align*}
where we defined $R_{r,i}(\mu):= \sum_{0\leq j_{1}\leq\cdots\leq j_{i}\leq r-i}\prod_{1\leq l\leq i}(\mu+j_{l}(1-\sigma)+l-1)$. The proof is completed by noting that, by means of simple algebraic manipulations, $R_{r+1,i}(\mu) = R_{r,i}(\mu+1-\sigma) + \mu R_{r,i-1}(\mu+1)$. Note that when $\nu_0(A) = 0$ and $i=r$, the convention $\nu_0(A)^{r-i} = 0^0=1$ is adopted. \hfill\qed

\medskip

\textsc{Proof of Proposition 1.} Let us consider the Borel sets $A_{0}:=\mathbb{X}\setminus\{X_{1}^{\ast},\ldots,X_{K_{n}}^{\ast}\}$ and $A_{l}:=\{X_{i}^{\ast}:N_{i,n}=l\}$, for any $l=1,\ldots,n$. The two parameter PD prior is a Gibbs-type prior with $h(t)=p(t;\sigma,\theta):=\sigma\Gamma(\theta)t^{-\theta}/\Gamma(\theta/\sigma)$, for any $\sigma\in(0,1)$ and $\theta>-\sigma$. Therefore one has $V_{n,k_{n}}=V_{p,(n,k_{n})}=[(\theta)_{n}]^{-1}\prod_{0\leq i\leq k_{n}-1}(\theta+i\sigma)$. By a direct application of Theorem 1 we can write
\begin{align*}
\E[Q^{r}_{h}(A_{0})\,|\,\boldsymbol{X}_{n}]&=\sum_{i=0}^{r}{r\choose i}(-1)^{i}\frac{(\theta)_{n}}{(\theta)_{n+i}}(n-\sigma k_{n})_{i}\\
&=(\theta)_{n}\frac{(\theta+\sigma k_{n})_{r}}{(\theta)_{n}(\theta+n)_{r}}\\
&=\frac{(\theta+\sigma k_{n})_{r}}{(\theta+ \sigma k_{n} +n-\sigma k_{n})_{r}},
\end{align*}
which is $r$-th moment of a Beta random variable with parameter $(\theta+\sigma k,n-\sigma k)$. Let us define the random variable $Y=Z_p R_{\sigma ,Z_p}$. Then, it can be easily verified that $Y$ has density function
\begin{align*}
f_Y(y) &= \int_{0}^{\infty} \frac{1}{z} f_{R_{\sigma,z}}(y/z)  f_{Z_p}(z)\ddr z \\
&=\frac{\sigma}{\Gamma(\theta/\sigma+k_{n})} \int_{0}^{\infty} \edr^{z^\sigma-y-z^\sigma} z^{\theta+\sigma k_{n}-2}f_\sigma(y/z) \ddr z\\
&=\frac{\sigma}{\Gamma(\theta/\sigma+k_{n})}  y^{\theta+\sigma k_{n}-1}  \edr^{-y} \int_{0}^{\infty} u^{-(\theta+\sigma k_n)} f_\sigma(u)  \ddr u
\end{align*}
where, by Equation 60 in \citet{Pit(03)}, $\int_{0}^{\infty} u^{-(\theta+\sigma k_n)} f_\sigma(u)  \ddr u =\Gamma(\theta/\sigma+k_n)/\sigma \Gamma(\theta+\sigma k_n)$. Hence $Y$ is a Gamma random variable with parameter $(\theta+\sigma k_{n},1)$. Accordingly, we have $W_{n-\sigma k_{n},Z_{p}}\stackrel{\text{d}}{=}B_{\theta+\sigma k_{n},n-\sigma k_{n}}$. Similarly, by a direct application of Theorem 1, for any $l>1$ we can write
\begin{align*}
\E[Q^{r}_{h}(A_{l})\,|\,\boldsymbol{X}_{n}]&=\frac{(\theta)_{n}}{(\theta)_{n+r}}((l-\sigma)m_{l,n})_{r}\\
&=\frac{((l-\sigma)m_{l,n})_{r}}{((l-\sigma)m_{l,n})_{r}+\theta+n-(l-\sigma)m_{l,n}},
\end{align*}
which is the $r$-th moment of a Beta random variable with parameter $((l-\sigma)m_{l,n},\theta+n-(l-\sigma)m_{l,n})$. Finally, the decomposition $B_{(l-\sigma)m_{l,n},\theta+n-(l-\sigma)m_{l,n}}\stackrel{\text{d}}{=}B_{(l-\sigma)m_{l,n},n-\sigma k_{n}-(l-\sigma)m_{l,n}}(1-W_{n-\sigma k_{n},Z_{p}})$ follows from a characterization of Beta random variables in Theorem 1 in \citet{Jam(54)}. It can be also easily verified by using the moments of Beta random variables.\hfill\qed

\medskip

\textsc{Proof of Proposition 2.} Let us consider the Borel sets $A_{0}:=\mathbb{X}\setminus\{X_{1}^{\ast},\ldots,X_{K_{n}}^{\ast}\}$ and $A_{l}:=\{X_{i}^{\ast}:N_{i,n}=l\}$, for any $l=1,\ldots,n$. The two parameter PD prior is a Gibbs-type prior with $h(t)=g(t;\sigma,\tau):=\exp\{\tau^{\sigma}-\tau t\}$, for any $\tau>0$. By a direct application of Theorem 1 we can write
\begin{align}\label{moment_gg}
&\E[Q^{r}_{g}(A_{0})\,|\,\boldsymbol{X}_{n}]\\
&\notag\quad\quad=\frac{\sigma\Gamma(n)}{C_{\sigma,\tau,n,{k_n}}\Gamma(n-\sigma {k_n})}\int_{0}^{1}w^{r}(1-w)^{n-1-\sigma {k_n}}\int_{0}^{+\infty}t^{-\sigma {k_n}}\edr^{-\tau t} f_{\sigma}(w t)\ddr t \ddr w,
\end{align}
where
\begin{align*}
C_{\sigma,\tau,n,{k_n}}&:=\frac{\sigma\Gamma(n)}{\Gamma(n-\sigma {k_n})}\int_{0}^{+\infty}t^{-\sigma {k_n}}\edr^{-\tau t}\int_{0}^{1}(1-w)^{n-1-\sigma {k_n}}f_{\sigma}(w t)\ddr w \ddr t\\
&=\sum_{i=0}^{n-1}{n-1\choose i}(-\tau)^{i}\Gamma(k-i/\sigma;\tau^{\sigma}).
\end{align*}
Hereafter we show that \eqref{moment_gg} coincides with the $r$-th moment of the random variable $W_{n-\sigma k_{n},Z_{g}}$. Given $Z_g=z$ it is easy to find that the distribution of $W_{n-\sigma k_{n},z}$ has the following density function
\begin{equation*}\label{cond_dens}
f_{W_{n-\sigma k_{n},z}}(w) =  \frac{\exp\{z^{\sigma}\}}{z\Gamma(n-{k_n}\sigma)}(1-w)^{n-{k_n}\sigma-1} \int_{0}^{+\infty}u^{n-{k_n}\sigma}\text{e}^{-u}f_{\sigma}\left(\frac{u w}{z}\right)\ddr u.
\end{equation*}
By randomizing over $z$ with respect to the distribution of $Z_g$ provides the distribution of $W_{n-\sigma k_{n},Z_{g}}$. Specifically,
\begin{align*}
f_{W_{n-\sigma k_{n},Z_{g}}}(w)&=\frac{\sigma}{C_{\sigma,\tau,n,{k_n}}\Gamma(n-\sigma k_{n})}(1-w)^{n-\sigma k_{n}-1}\\
&\quad\times\int_{\tau}^{\infty} z^{-n+\sigma k_{n}-1}(z-\tau)^{n-1}\int_0^{\infty} u^{n-\sigma k_{n}} \edr^{-u} f_\sigma\left(\frac{u w}{z}\right)\ddr u \ddr z\\
&=\frac{\sigma}{C_{\sigma,\tau,n,k_n}\Gamma(n-\sigma k)}(1-w)^{n-\sigma k_{n}-1}\\
&\quad\times\int_{\tau}^{\infty} (z-\tau)^{n-1}\int_0^{\infty} t^{n-\sigma k_{n}} \edr^{-tz} f_\sigma\left(w t\right)\ddr t \ddr z\\
&=\frac{\sigma\Gamma(n)}{C_{\sigma,\tau,n,k_n}\Gamma(n-\sigma k_{n})}(1-w)^{n-\sigma k_{n}-1}\int_{0}^{\infty} t^{-\sigma k_{n}} \edr^{-\tau t} f_\sigma\left(w t\right)\ddr t. 
\end{align*}
Therefore,
\begin{align*}
&\E[W_{n-\sigma k_{n},Z_{g}}^r]\\
&\quad=\frac{\sigma\Gamma(n)}{C_{\sigma,\tau,n,k_n}\Gamma(n-\sigma k_{n})}\int_0^1 w^r (1-w)^{n-\sigma k_{n}-1}\int_{0}^{\infty} t^{-\sigma k_{n}} \edr^{-\tau t} f_\sigma\left(w t\right)\ddr t \ddr w
\end{align*}
which coincides with \eqref{moment_gg}. We complete the proof by determining the distribution of the random variable $Q_{g}(A_{l})\,|\,\boldsymbol{X}_{n}$, for any $l>1$. Again, by a direct application of Theorem 1 we can write
\begin{align*}
&\E[Q_{g}^{r}(A_{l})\,|\,\boldsymbol{X}_{n}]\\
&\quad=((l-\sigma)m_{l,n})_{r}\frac{\frac{\sigma^{k_{n}}}{\Gamma(n-\sigma k_{n}+r)}}{\frac{\sigma^{k_{n}}}{\Gamma(n-\sigma k_{n})}}\frac{\int_{0}^{+\infty}t^{-\sigma k_{n}}\exp\{-\tau t\}\int_{0}^{1}(1-z)^{n+r-1-\sigma k_{n}}f_{\sigma}(zt)\ddr t \ddr z}{\int_{0}^{+\infty}t^{-\sigma k_{n}}\exp\{-\tau t\}\int_{0}^{1}(1-z)^{n-1-\sigma k_{n}}f_{\sigma}(zt)\ddr t \ddr z}\\
&\quad=\frac{\Gamma(n-\sigma k_{n})}{\Gamma\left((l-\sigma)m_{l,n}\right)\Gamma(\sum_{1\leq i\neq l\leq n }im_{i,n}-\sigma\sum_{1\leq i\neq l\leq n }m_{i,n})}\\
&\quad\quad\times\int_{0}^{1}x^{(l-\sigma)m_{l,n}+r-1}(1-x)^{\sum_{1\leq i\neq l\leq n }im_{i,n}-\sigma\sum_{1\leq i\neq l\leq n }m_{i,n}-1}\\
&\quad\quad\quad\times\frac{\int_{0}^{+\infty}t^{-\sigma k_{n}}\exp\{-\tau t\}\int_{0}^{1}(1-z)^{n+r-1-\sigma k_{n}}f_{\sigma}(zt)\ddr t \ddr z}{\int_{0}^{+\infty}t^{-\sigma k_{n}}\exp\{-\tau t\}\int_{0}^{1}(1-z)^{n-1-\sigma k_{n}}f_{\sigma}(zt)\ddr t \ddr z}\ddr x\\
&\quad=\frac{\Gamma(n-\sigma k_{n})}{\Gamma\left((l-\sigma)m_{l,n}\right)\Gamma(\sum_{1\leq i\neq l\leq n }im_{i,n}-\sigma\sum_{1\leq i\neq l\leq n }m_{i,n})}\\
&\quad\quad\times\int_{0}^{1}x^{(l-\sigma)m_{l,n}-1}(1-x)^{\sum_{1\leq i\neq l\leq n }im_{i,n}-\sigma\sum_{1\leq i\neq l\leq n }m_{i,n}-1}\\
&\quad\quad\quad\times\frac{\frac{\sigma\Gamma(n)}{\Gamma(n-\sigma k_{n})}\int_{0}^{+\infty}t^{-\sigma k_{n}}\exp\{-\tau t\}\int_{0}^{1}x^{r}(1-z)^{r}(1-z)^{n-1-\sigma k_{n}}f_{\sigma}(zt)\ddr t \ddr z}{\frac{\sigma^{k_{n}}}{\Gamma(n-\sigma k_{n})}\int_{0}^{+\infty}t^{-\sigma k_{n}}\exp\{-\tau t\}\int_{0}^{1}(1-z)^{n-1-\sigma k_{n}}f_{\sigma}(zt)\ddr t \ddr z}\ddr x,
\end{align*}
which is the $r$-th moment of the scale mixture $B_{(l-\sigma)m_{l,n},n-\sigma k_{n}-(l-\sigma)m_{l,n}}(1-W_{n-\sigma k_{n},Z_{g}})$, where $W_{n-\sigma k_{n},Z_{g}}$ is the random variable characterized above, and where the Beta random variable $B_{(l-\sigma)m_{l,n},n-\sigma k_{n}-(l-\sigma)m_{l,n}}$ is independent of the random variable $(1-W_{n-\sigma k_{n},Z_{g}})$. The proof is completed. \hfill\qed

\medskip

\textsc{Proof of Theorem 2.} According to the fluctuation limit \eqref{eq:asymp_dist} there exists a nonnegative and finite random variable $S_{\sigma,h}$ such that $n^{-\sigma}K_{n}\stackrel{\text{a.s.}}{\longrightarrow}S_{\sigma,h}$ as $n\rightarrow+\infty$. Let $\Omega_{0}:=\{\omega\in\Omega : \lim_{n\rightarrow+\infty}n^{-\sigma}K_{n}(w)=S_{\sigma,h}(\omega)\}$. Furthermore, let us define $g_{0,h}(n,k_{n})=V_{h,(n+1,k_{n}+1)}/V_{h,(n,k_{n})}$, where $V_{h,(n,k_{n})}=\sigma^{k_{n}-1}\Gamma(k_{n})\E[h(S_{\sigma,k_{n}}/B_{\sigma k_{n},n-\sigma k_{n}})]/\Gamma(n)$. Then we can write the following expression
\begin{equation}\label{eq:ratio}
g_{0,h}(n,k_n) = \frac{\sigma k_n}{n}\frac{\E\left[h\left(\frac{S_{\sigma,k_{n}+1}}{B_{\sigma k_{n}+1,n+1-\sigma (k_{n}+1)}}\right)\right]}{\E\left[h\left(\frac{S_{\sigma,k_{n}}}{B_{\sigma k_{n},n-\sigma k_{n}}}\right)\right]}.
\end{equation}
We have to show that the ratio of the expectations in \eqref{eq:ratio} converges to $1$ as $n\rightarrow+\infty$. For this, it is sufficient to show that, as $n\rightarrow+\infty$, the random variable $T_{\sigma,n,k_n}=S_{\sigma,k_{n}}/B_{\sigma k_{n},n-\sigma k_{n}}$ converges almost surely to a random variable $T_{\sigma,h}$. This is shown by computing the moment of order $r$ of $T_{\sigma,n,k_n}$, i.e.,
\begin{align*}
\E(T_{\sigma,n,k_n}^r) = \frac{\Gamma(n)}{\Gamma(n-r)}\frac{\Gamma(k_n-r/\sigma)}{\Gamma(k_n)}\simeq\frac{n^r}{k_n^{r/\sigma}}.
\end{align*}
For any $\omega\in\Omega_{0}$ the ratio $n/K^{1/\sigma}_n(\omega)=n/k_n^{1/\sigma}$ converges to $S_{\sigma,h}^{-1/\sigma}(\omega)=T_{\sigma,h}(\omega)=t$. Accordingly, $n^{r}/k_n^{r/\sigma}$ converges to $\E[T_{\sigma}^{r}(\omega)]=t^{r}$ for any $\omega\in\Omega_{0}$. Since $\P[\Omega_{0}]=1$, the almost sure limit, as $n$ tends to infinity, of the random variable $T_{\sigma,n,K_n}$ is identified with the nonnegative random variable $T_{\sigma,h}$, which has density function $f_{T_{\sigma,h}}(t)=h(t)f_{\sigma}(t)$. The proof is completed.

\medskip

\textsc{Proof of Proposition 3.} Let $h(t)=p(t;\sigma,\theta):=\sigma\Gamma(\theta)t^{-\theta}/\Gamma(\theta/\sigma)$, for any $\sigma\in(0,1)$ and $\theta>-\sigma$. Furthermore, let us define $g_{0,p}(n,k_{n})=V_{p,(n+1,k_{n}+1)}/V_{p,(n,k_{n})}$ and  $g_{1,p}(n,k_{n})=1-V_{p,(n+1,k_{n}+1)}/V_{p,(n,k_{n})}$, so that we have $g_{0}(n,k_{n})=(\theta+\sigma k_{n})/(\theta+n)$ and $g_{1}(n,k_{n})=1/(\theta+n)$. Then,
\begin{equation}\label{sec_pd_1}
g_{0,p}(n,k_n)=\frac{\sigma k_{n}}{n}+\frac{\theta}{n}+o\left(\frac{1}{n}\right)
\end{equation}
and
\begin{equation}\label{sec_pd_2}
g_{1,p}(n,k_n)=\frac{1}{n}-\frac{\theta}{n^{2}}+o\left(\frac{1}{n^{2}}\right)
\end{equation}
follow by a direct application of the Taylor series expansion to $g_{0}(n,k_{n})$ and $g_{1}(n,k_{n})$, respectively, and then truncating the series at the second order. The proof is completed by combining \eqref{sec_pd_1} and \eqref{sec_pd_2} with the Bayesian nonparametric estimator $\hat{\mathcal{D}}_{n}(l)$ under a two parameter PD prior.\hfill\qed

\medskip

\textsc{Proof of Proposition 4.} The proof is along lines similar to the proof of  Proposition 3.2. in \citet{Rug(13)}, which, however, considers a different parameterization for the normalized GG prior. Let $h(t)=g(t;\sigma,\tau):=\exp\{\tau^{\sigma}-\tau t\}$, for any $\sigma\in(0,1)$ and $\tau>0$, and let $g_{0,g}(n,k_{n})=V_{g,(n+1,k_{n}+1)}/V_{g,(n,k_{n})}$ and  $g_{1,p}(n,k_{n})=1-V_{g,(n+1,k_{n}+1)}/V_{g,(n,k_{n})}$, where we have
\begin{displaymath}
V_{g,(n,k_{n})}=\frac{\sigma^{k_n}\exp\{\tau ^{\sigma}\}}{\Gamma(n)}\int_{0}^{+\infty}x^{n-1}(\tau +x)^{-n+\sigma k_n}\text{e}^{-(\tau +x)^{\sigma}}\ddr x.
\end{displaymath}
Note that, by using the triangular relation characterizing the nonnegative weight $\Vgnk$, we can write
\begin{displaymath}
g_{0,g}(n,k_n) =\frac{\Vgnk-(n-\sigma k_n)\Vgnuk}{\Vgnk}=1-\left(1-\frac{\sigma k_n}{n}\right)w(n,k_n),
\end{displaymath}
where
\begin{displaymath}
w(n,k_n)=\frac{\int_{0}^{\infty}x^{n}\exp\{-[(\tau
+x)^{\sigma}-\tau^{\sigma}]\}(\tau+x)^{\sigma k_n-n-1}\,\ddr x}
{\int_{0}^{\infty}x^{n-1}\exp\{-[(\tau+x)^{\sigma
}-\tau
^{\sigma}]\}(\tau+x)^{\sigma k_n-n}\,\ddr x}.
\end{displaymath}
Let us denote by $f(x)$ the integrand function of the denominator of $1-w(n,k_n)$, and let  $f_{N}(x) = \tau f(x)/(\tau+x)$. That is, $f_{N}(x)$ is the denominator of $1-w(n,k_n)$. Therefore we can write
\begin{displaymath}
1-w(n,k_n)=\frac{\int_{0}^{\infty}\tau f(x)/(\tau+x)\,\ddr x}{\int_{0}^{\infty
}f(x)\,\ddr x}.
\end{displaymath}
Since $f(x)$ is unimodal, by means of the Laplace approximation method it can be approximated with a Gaussian kernel with mean $x^{*}=\argmax_{x>0} x^{n-1}\exp\{-[(\tau
+x)^{\sigma}-\tau^{\sigma}]\}(\tau+x)^{\sigma k_n-n}$ and with variance $-[(\log\circ f)''(x^*)]^{-1}$. The same holds for $f_{N}(x)$. Then, we obtain the approximation
\begin{displaymath}
1-w(n,k_n)\simeq\frac{f_{N}(x^{*}_{N})C(x^{*}_{N},-[(\log\circ f_N)''(x_N^*)]^{-1})}
{f(x^{*}_{D})C(x^{*}_{D},-[(\log\circ f)''(x_D^*)]^{-1})},
\end{displaymath}
where $x^{*}_{N}$ and $x^{*}_{D}$ denote the modes of $f_N$ and $f$, respectively, and where $C(x,y)$ denotes the normalizing constant of a Gaussian kernel with mean $x$ and variance $y$. Specifically, this yields to 
\begin{equation}\label{approx-phi}
1-w(n,k_n)\simeq
\frac{f_{N}(x^{*}_{N})}{f(x^{*}_{D})}
\left(\frac{(\log\circ f_N)''(x^{*}_{N})}{(\log\circ f)''(x^{*}_{D})}\right)^{-1/2}.
\end{equation}
The mode $x^{*}_{D}$ is the only positive real root of the function $G(x) = \sigma x(\tau+x)^{\sigma}-(n-1)\tau-(\sigma k_n-1)x$. A study of $G$ shows that $x^{*}_{D}$ is bounded by below by a positive constant times $n^{1/(1+\sigma)}$, which implies that the terms involving $\tau$ are negligible in the following renormalization of $G(x^{*}_{D})$ 
\begin{displaymath}
\sigma\frac{x^{*}_{D}}{n}\left(\frac{\tau}{n}+\frac{x^{*}_{D}}{n}\right)^\sigma-\frac{n-1}{n^{\sigma+1}}\tau-\frac{\sigma k_n-1}{n^\sigma}\frac{x^{*}_{D}}{n}.
\end{displaymath}
The same calculation holds for $x^{\ast}_{N}$. According to the fluctuation limit \eqref{eq:asymp_dist} there exists a nonnegative and finite random variable $S_{\sigma,g}$ such that $n^{-\sigma}K_{n}\stackrel{\text{a.s.}}{\longrightarrow}S_{\sigma,g}$ as $n\rightarrow+\infty$. Let $\Omega_{0}:=\{\omega\in\Omega : \lim_{n\rightarrow+\infty}n^{-\sigma}K_{n}(w)=S_{\sigma,h}(\omega)\}$, and let $S_{\sigma,g}(\omega)=s_{\sigma}$ for any $\omega\in\Omega_{0}$. Then, we have
\begin{equation}\label{eq:equiv-x-star}
\frac{x_N^*}{n}\simeq \frac{x_D^*}{n}\simeq s_\sigma^{1/\sigma}.
\end{equation}
In order to make use of  \eqref{approx-phi}, we also need  an asymptotic equivalence for $x_D^*-x_N^*$. Note that $G(x_D^*)=0$ and $G(x_N^*)=-x_N^*$ allow us to resort to a first order Taylor bound on $G$ at $x_N^*$ and shows that $x_D^*-x_N^*$ has a lower bound equivalent to $s_\sigma^{(1-\sigma)/\sigma}n^{1-\sigma}/\sigma^2$. The same argument applied to $G(x)+x$ at $x_D^*$ provides an upper bound with the same asymptotic equivalence, thus
\begin{equation}\label{eq:equiv-diff-x}
\frac{x_D^*-x_N^*}{n^{1-\sigma}}\simeq \frac{s_\sigma^{(1-\sigma)/\sigma}}{\sigma^2}.
\end{equation}
By studying $f$ and $f_N$, as well as the second derivative of their logarithm, together with asymptotic equivalences \eqref{eq:equiv-x-star} and \eqref{eq:equiv-diff-x}, we can write  $f(x_D^*) \simeq f(x_N^*)$ and $(\log\circ f)''(x_D^*) \simeq (\log\circ f)''(x_N^*) \simeq (\log\circ f_N)''(x_N^*)$. Hence, from \eqref{approx-phi} one obtains $1-w(n,k_n) \simeq \tau/(\tau+x_N^*) \simeq \tau s_\sigma^{-1/\sigma}/n$, which leads to
\begin{align}\label{sec_gg_1}
\notag g_{0,g}(n,k_n) &=1 - \left(1-\frac{\sigma k_n}{n}\right)\left(1-\tau s_\sigma^{-1/\sigma}\frac{1}{n} +o\left(\frac{1}{n}\right)\right),\\
&=\frac{\sigma k_n}{n}+\tau s_\sigma^{-1/\sigma}\frac{1}{n}+ o\left(\frac{1}{n}\right),
\end{align}
and
\begin{align}\label{sec_gg_2}
\notag g_{1,g}(n,k_n)&=\frac{1-g_{0,g}(n,k_n)}{n-\sigma k_n} = \frac{1}{n}\left(1-\frac{\tau s_\sigma^{-1/\sigma}/n +\smallon}{1-\frac{\sigma k}{n}}\right),\\
&=\frac{1}{n}\left(1-\frac{\tau s_\sigma^{-1/\sigma}}{n} +\smallon\right).
\end{align}
Expressions \eqref{sec_gg_1} and \eqref{sec_gg_2} provide second order approximations of $ g_{0,g}(n,k_n)$ and $ g_{1,g}(n,k_n)$, respectively. Recall that for any $\omega$ in $\Omega_{0}$ we have $n^{-\sigma}k_n \simeq s_\sigma$, namely we can replace $s_{\sigma}$ with $n^{-\sigma}k_n$. This is because of the fluctuation limit displayed in \eqref{eq:asymp_dist}. The proof is completed by combining \eqref{sec_gg_1} and \eqref{sec_gg_2} with the Bayesian nonparametric estimator $\hat{\mathcal{D}}_{n}(l)$ under a normalized GG prior.\hfill\qed


\section{Details on the derivation of $\hat{\mathcal{D}}_{n}(l)\simeq \check{\mathcal{D}}_{n}(l;\mathscr{S}_{\text{PD}})$}
\setcounter{equation}{0}

Let us define $c_{\sigma,l}=\sigma(1-\sigma)_{l-1}/l!$ and recall that $\hat{\mathcal{D}}_{n}(0)=V_{n+1,k_{n}+1}/V_{n,k_{n}}$ and $\hat{\mathcal{D}}_{n}(l)=(l-\sigma)m_{l,n}V_{n+1,k_{n}}/V_{n,k_{n}}$. The relationship between the Bayesian nonparametric estimator $\hat{\mathcal{D}}_{n}(l)$ and the smoothed Good-Turing estimator $\check{\mathcal{D}}_{n}(l;\mathscr{S}_{\text{PD}})$ follows by combining Theorem 2 with the fluctuation limits \eqref{eq:asymp_dist} and \eqref{eq:asymp_dist_freq}. For any $\omega\in\Omega$, a version of the predictive distributions of $Q_{\sigma,h}$ is
\begin{displaymath}
\frac{V_{n+1,K_{n}(\omega)+1}}{V_{n,K_{n}(\omega)}}\nu_{0}(\cdot)+\frac{V_{n+1,K_{n}(\omega)}}{V_{n,K_{n}(\omega)}}\sum_{i=1}^{K_{n}(\omega)}(N_{i,n}(\omega)-\sigma)\delta_{X_{i}^{\ast}(\omega)}(\cdot).
\end{displaymath} 
According to \eqref{eq:asymp_dist} and \eqref{eq:asymp_dist_freq}, $\lim_{n\rightarrow+\infty}c_{\sigma,l}M_{l,n}/K_{n}=1$ almost surely. See Lemma 3.11 in \citet{Pit(06)} for additional details. By Theorem 2 we have $V_{n+1,K_{n}+1}/V_{n,K_{n}}\stackrel{\text{a.s.}}{\simeq}\sigma K_{n}/n$, and $M_{1,n}\stackrel{\text{a.s.}}{\simeq}\sigma K_{n}$, as $n\rightarrow+\infty$. Then, a version of the Bayesian nonparametric estimator of the $0$-discovery coincides with
\begin{align}\label{eq_vers_stim1}
\frac{V_{n+1,K_{n}(\omega)+1}}{V_{n,K_{n}(\omega)}}&\simeq\frac{\sigma K_{n}(\omega)}{n}\\
&\notag\simeq\frac{M_{1,n}(\omega)}{n},
\end{align}
as $n\rightarrow+\infty$. By Theorem 2 we have $V_{n+1,K_{n}}/V_{n,K_{n}}\stackrel{\text{a.s.}}{\simeq}1/n$, and $M_{l,n}\stackrel{\text{a.s.}}{\simeq}c_{\sigma,l} K_{n}$, as $n\rightarrow+\infty$. Accordingly, a version of the Bayesian nonparametric estimator of the $l$-discovery coincides with
\begin{align}\label{eq_vers_stim2}
(l-\sigma)M_{l,n}(\omega)\frac{V_{n+1,K_{n}(\omega)}}{V_{n,K_{n}(\omega)}}&\simeq(l-\sigma)\frac{M_{l,n}(\omega)}{n}\\
&\notag\simeq c_{\sigma,l}(l-\sigma)\frac{K_{n}(\omega)}{n}\\
&\notag\simeq(l+1)\frac{M_{l+1,n}(\omega)}{n},
\end{align}
as $n\rightarrow+\infty$. Let $\Omega_{0}:=\{\omega\in\Omega : \lim_{n\rightarrow+\infty}n^{-\sigma}K_{n}(w)=Z_{\sigma,\theta/\sigma}(\omega),\lim_{n\rightarrow+\infty}n^{-\sigma}M_{l,n}(\omega)=c_{\sigma,l}Z_{\sigma,\theta/\sigma}(\omega)\}$. From \eqref{eq:asymp_dist} and \eqref{eq:asymp_dist_freq} we have $\P[\Omega_{0}]=1$. Fix $\omega\in\Omega_{0}$ and denote by $k_{n}=K_{n}(\omega)$ and $m_{l,n}=M_{l,n}(\omega)$ the number of species generated and the number of species with frequency $l$ generated by the sample $\boldsymbol{X}_{n}(\omega)$. Accordingly, $\hat{\mathcal{D}}_{n}(l)\simeq \check{\mathcal{D}}_{n}(l;\mathscr{S}_{\text{PD}})$ follows from \eqref{eq_vers_stim1} and \eqref{eq_vers_stim2}.


\section{Additional illustrations}
\setcounter{equation}{0}

In this Section we provide additional illustrations accompanying those of Section 4 in the main manuscript. Specifically, we consider a Zeta distribution with parameter $s=1.5$. We draw 500 samples of size $n=1000$ from such distribution, we order them according to the number of observed species $k_n$, and we split them in 5 groups: for $i=1,2,\ldots,5$, the $i$-th group of samples will be composed by 100 samples featuring a total number of observed species $k_n$ that stays between the quantiles of order $(i-1)/5$ and $i/5$ of the empirical distribution of $k_n$. Then we pick at random one sample for each group and label it with the corresponding index $i$. This procedure leads to five samples. As shown in Table~S1, the choice of $s=1.5$ leads to samples with a smaller number of distinct values if compared with the case $s=1.1$ (see also Table~1 in the main manuscript).  Table~S2, under the two parameter PD prior and the
normalized GG prior, shows the estimated $l$-discoveries, for $l=0,1,5,10$, and the corresponding 95\% posterior credible intervals.
Finally, Figure S1 shows how the average ratio $\bar{r}_{1,2,n}$ evolves as the sample size increases (see Section 4.2 in the main manuscript).

\newpage

\noindent Table S1: Simulated data with $s=1.5$. For each sample we report the sample size $n$, the number of species ${k_n}$ and the maximum likelihood values $(\hat \sigma, \hat \theta)$ and $(\hat \sigma, \hat \tau)$.
   \begin{small}
      \begin{center}
      \begin{tabular}{|cccc|cc|cc|}
      \hline
         \multicolumn{4}{|c|}{} &  \multicolumn{2}{c|}{PD} & \multicolumn{2}{c|}{GG} \\
      \hline
      & sample & $n$ & ${k_n}$ & $\hat\sigma$ & $\hat\theta$ & $\hat \sigma$ & $\hat\tau$ \\
    \hline
          \multirow{5}{*}{Simulated data } & $ 1$ & 1000  & 128  &     0.624  &  1.207 & 0.622 &3.106\\
      & $ 2$ & 1000  &   135  &     0.675  & 0.565 &     0.673 &    0.957\\
      & $ 3$ & 1000  &  138  &    0.684  &  0.487 &    0.682  &    0.795\\      
      & $ 4$ & 1000  &  146 & 0.656  &  1.072 &    0.655  &    2.302\\
      & $ 5$ &  1000  & 149  &  0.706  &  0.377 &    0.704  &    0.592\\    
    \hline 
      \end{tabular}
      \end{center}
 \end{small}\bigskip

\noindent Table S2: Simulated data with $s=1.5$. We report the true value of the probability $D_{n}(l)$ and the Bayesian nonparametric estimates of $D_{n}(l)$ with 95\% credible intervals.
     \begin{small}
      \begin{center}
      \begin{tabular}{|ccc|cc|cc|cc|}
      \hline 
         \multicolumn{3}{|c|}{} &  \multicolumn{2}{c|}{Good--Turing} &   \multicolumn{2}{c|}{PD} & \multicolumn{2}{c|}{GG} \\
      \hline
       $l$ & sample & $D_{n}(l)$ & $\check{\mathcal{D}}_n(l)$ & 95\%-c.i. & $\hat{\mathcal{D}}_n(l)$ & 95\%-c.i. & $\hat{\mathcal{D}}_n(l)$ & 95\%-c.i.  \\
    \hline
      \multirow{5}{*}{$0$}
      & $1$ & 0.099 & 0.080 & (0.010, 0.150) & 0.081 & (0.065, 0.098)  & 0.081 & (0.065, 0.098) \\
      & $2$ & 0.103 & 0.092 & (0.012, 0.172) & 0.092 & (0.075, 0.110)  & 0.091 & (0.075, 0.110) \\
      & $3$ & 0.095 & 0.096 & (0.014, 0.178) & 0.095 & (0.078, 0.114)  & 0.095 & (0.076, 0.113) \\
      & $4$ & 0.096 & 0.096 & (0.015, 0.177) & 0.097 & (0.079, 0.116)  & 0.097 & (0.080, 0.115) \\
      & $5$ & 0.093 & 0.108 & (0.019, 0.197) & 0.106 & (0.087, 0.126)  & 0.105 & (0.087, 0.124) \\
     \hline      
          \multirow{5}{*}{$1$} 
      & $1$ & 0.030 & 0.038 & (0.031, 0.045 ) & 0.030 & (0.020, 0.042)  & 0.030 & (0.021, 0.042) \\
      & $2$ & 0.037 & 0.030 & (0.024, 0.036) & 0.030 & (0.021, 0.041)  & 0.030 & (0.020, 0.042) \\
      & $3$ & 0.034 & 0.034 & (0.028, 0.040) & 0.030 & (0.021, 0.042)  & 0.031 & (0.021, 0.042) \\
      & $4$ & 0.029 & 0.040 & (0.033, 0.047) & 0.033 & (0.023, 0.045)  & 0.033 & (0.022, 0.044) \\
      & $5$ & 0.040 & 0.026 & (0.021, 0.031) & 0.032 & (0.022, 0.044)  & 0.032 & (0.023, 0.043) \\
    \hline 
              \multirow{5}{*}{$5$} 
      & $1$ & 0.013 & 0.012 & (0.008, 0.016) & 0.013 & (0.007, 0.021)  & 0.013 & (0.007, 0.021) \\
      & $2$ & 0.011 & 0.006 & (0.003, 0.009) & 0.004 & (0.001, 0.009)  & 0.004 & (0.001, 0.009) \\
      & $3$ & 0.010 & 0.012 & (0.007, 0.017) & 0.009 & (0.004, 0.015)  & 0.009 & (0.004, 0.016) \\
      & $4$ & 0.010 & 0.036 & (0.024, 0.048) & 0.009 & (0.004, 0.015)  & 0.009 & (0.004, 0.015) \\
      & $5$ & 0.012 & 0 & (0, 0) & 0.013 & (0.007, 0.021)  & 0.013 & (0.006, 0.021) \\
    \hline 
              \multirow{5}{*}{$10$} 
      & $1$ & 0.019 & 0 & (0, 0) & 0.019 & (0.011, 0.028)  & 0.019 & (0.011, 0.028) \\
      & $2$ & 0 & 0.011 & n.a. & 0 & (0, 0)  & 0 & (0,0) \\
      & $3$ & 0.011 & 0.011 & (0.006, 0.016) & 0.009 & (0.004, 0.016)  & 0.009 & (0.004, 0.016) \\
      & $4$ & 0 & 0 & n.a. & 0 & (0,0)  & 0 & (0,0) \\
      & $5$ & 0.006 & 0 & (0, 0) & 0.009 & (0.004, 0.016)  & 0.009 & (0.004, 0.017) \\
    \hline
      \end{tabular}
      \end{center}
   \end{small}

\newpage
\begin{center}
\includegraphics[width=0.6\linewidth]{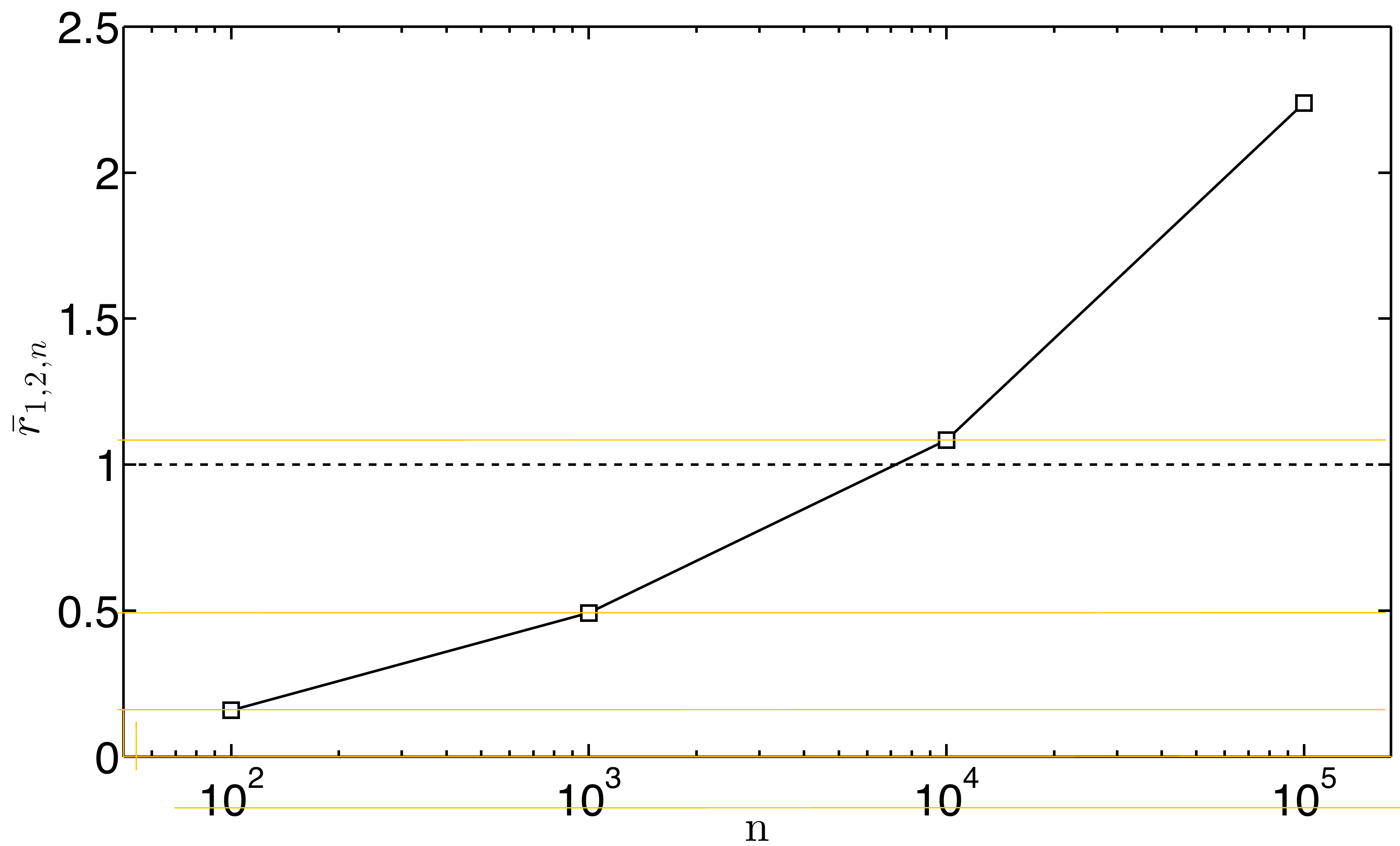}
\end{center}

\noindent Figure S1: Average ratio $\bar{r}_{1,2,n}$ of sums of squared approximation errors for different sample sizes $n=10^2,10^3, 10^4,10^5$. For the $x$-axis a logarithmic scale was used.


\section*{Acknowledgements}

The authors are grateful to two anonymous referees for valuable comments and suggestions, and to Alexander Gnedin for suggesting the asymptotic relationship \eqref{eq:Gnedin_refinement}. Special thanks are due to Stefano Peluchetti (\url{http://www.scilua.org/}) for numerous useful discussions on numerical optimization in Lua. BN would like to thank Collegio Carlo Alberto where he was working when part of the research presented here was carried out. JA and SF are supported by the European Research Council through StG N-BNP 306406. YWT is supported by the European Research Council through the European Unions Seventh Framework Programme (FP7/2007-2013) ERC grant agreement 617411.
\par


\bibhang=1.7pc
\bibsep=2pt
\fontsize{9}{14pt plus.8pt minus .6pt}\selectfont
\renewcommand\bibname


{\large \bf References}
\begin{thebibliography}{11}
\expandafter\ifx\csname
natexlab\endcsname\relax\def\natexlab#1{#1}\fi
\expandafter\ifx\csname url\endcsname\relax
  \def\url#1{\texttt{#1}}\fi
\expandafter\ifx\csname urlprefix\endcsname\relax\def\urlprefix{URL
}\fi

\bibitem[Arbel et al.(2016)]{arbel2014full} 
Arbel, J., Lijoi, A., and Nipoti, B. (2016). Full Bayesian inference with hazard mixture models. \textit{Comput. Statist. Data Anal.}. \textbf{93}, 359--372.

\bibitem[Baayen(2001)]{Baa(01)}
Baayen, R. H. (2001). \textit{Word Frequency Distributions}. Springer Science and Business Media.

\bibitem[Bubeck et al.(2013)]{Bub(13)} 
Bubeck, S., Ernst, D., and Garivier, A. (2013). Optimal discovery with probabilistic expert advice: finite time analysis and macroscopic optimality, \textit{J. Mach. Learn. Res.}, \textbf{14}, 601--623.

\bibitem[Bunge and Fitzpatrick(1993)]{Bun(93)} 
Bunge, J. and Fitzpatrick, M. (1993). Estimating the number of species: a review. \textit{J. Amer. Statist. Assoc.}. \textbf{88}, 364--373.

\bibitem[Bunge et al.(2014)]{Bun(14)} 
Bunge, J., Willis, A., and Walsh, F. (2014). Estimating the number of species in microbial diversity studies. \textit{Annu. Rev. Sta. Appl.}. \textbf{1}, 427--445.

\bibitem[Caron and Fox(2015)]{Car(15)} 
Caron, F. and Fox, E.B. (2015). Sparse graphs with exchangeable random measures. \textit{Preprint ArXiv:1401.1137}.

\bibitem[Chao and Lee(1992)]{Cha(92)} 
Chao, A. and Lee, S. (1992). Estimating the number of classes via sample coverage. \textit{J. Amer. Statist. Assoc.}, \textbf{87}, 210--217.

\bibitem[Chao et al.(2009)]{Cha(09)} 
Chao, A., Colwell, R.K., Lin, C.W. and Gotelli, N.J. (2009). Sufficient sampling for asymptotic minimum species richness estimators. \textit{Ecology}. \textbf{90}, 1125--1133.

\bibitem[Church and Gale(1991)]{Chu(91)} 
Church, K.W. and Gale, W.A. (1991). A comparison of the enhanced Good--Turing and delated estimation methods for estimating probabilities of english bigrams. \textit{Comput. Speech Lang.}. \textbf{5}, 19--54.

\bibitem[De Blasi et al.(2015)]{Deb(15)} 
De Blasi, P., Favaro, S., Lijoi, A., Mena, R.H., Pr\"unster, I., and Ruggiero, M. (2015). Are Gibbs-type priors the most natural generalization of the Dirichlet process? \textit{IEEE Trans. Pattern Anal. Mach. Intell.}. \textbf{37}, 212--229.

\bibitem[Devroye(2009)]{Dev(09)}
Devroye, L. (2009). Random variate generation for exponentially and polynomially tilted stable distributions. \textit{ACM Trans. Model. Comput. Simul.}. \textbf{4}: 18.

\bibitem[Efron and Thisted(1976)]{Efr(76)}  
Efron, B. and Thisted, R. (1976). Estimating the number of unseen species: How many words did Shakespeare know? \textit{Biometrika}. \textbf{63}, 435--447.

\bibitem[Engen(1978)]{Eng(78)}  
Engen, S. (1978). \textit{Stochastic Abundance Models.} Chapman and Hall.

\bibitem[Favaro et~al., 2009]{favaro2009bayesian}
Favaro, S., Lijoi, A., Mena, R.~H., and Pr{\"u}nster, I. (2009).
\newblock {Bayesian non-parametric inference for species variety with a
  two-parameter Poisson--Dirichlet process prior}.
\newblock {\em J. Roy. Stat. Soc. B}. \textbf{71},993--1008.

\bibitem[Favaro et al.(2012)]{Fav(12)} 
Favaro, S., Lijoi, A., and Pr\"unster, I. (2012). A new estimator of the discovery probability. \textit{Biometrics}. \textbf{68}, 1188--1196.

\bibitem[Favaro et al.(2016)]{Fav(15)} 
Favaro, S., Nipoti, B., and Teh, Y.W. (2016). Rediscovery Good--Turing estimators via Bayesian nonparametrics. \textit{Biometrics}. \textbf{72}, 136--145.

\bibitem[Ferguson(1973)]{Fer(73)} 
Ferguson, T.S. (1973). A Bayesian analysis of some nonparametric problems. \textit{Ann. Statist.}. \textbf{1}, 209--230.

\bibitem[Gilks and Wild(1992)]{Gil(92)}
Gilks, W.R. and Wild, P. (1992). Adaptive rejection sampling for Gibbs sampling. \textit{Appl. Statist.}. \textbf{41}, 337--348.

\bibitem[Gnedin et al.(2007)]{Gne(07)}
Gnedin, A., Hansen, B., and Pitman, J. (2007). Notes on the occupancy problem with infinitely many boxes: general asymptotics and power law. \textit{Probab. Surv.}. \textbf{4}, 146--171.

\bibitem[Gnedin and Pitman(2006)]{Gne(06)}
Gnedin, A. and Pitman, J. (2005). Exchangeable Gibbs partitions and Stirling triangles. \textit{J. Math. Sci.}. \textbf{138}, 5674--5685.

\bibitem[Good(1953)]{Goo(53)}  
Good, I.J. (1953). The population frequencies of species and the estimation of population parameters. \textit{Biometrika}. \textbf{40}, 237--64.

\bibitem[Guindani et al.(2014)]{Gui(14)}
Guindani, M., Sepulveda, N., Paulino, C.D. and M\"uller, P. (2014). A Bayesian semiparametric approach for the differential analysis of sequence data. \textit{J. Roy. Stat. Soc. C}. \textbf{63}, 385--404.

\bibitem[Jambunathan(1954)]{Jam(54)}
Jambunathan, M.V. (1954). Some Properties of Beta and Gamma Distributions. \textit{Ann. Math. Statist.}. \textbf{25}, 401--405.

\bibitem[James(2002)]{Jam(02)}
James, L.F. (2002). Poisson process partition calculus with applications to exchangeable models and Bayesian nonparametrics. \textit{Preprint  arXiv:math/0205093}.

\bibitem[Lijoi et al.(2007)]{Lij(07)}   
Lijoi, A., Mena, R.H., and Pr\"unster, I. (2007). Bayesian nonparametric estimation of the probability of discovering new species. \textit{Biometrika}. \textbf{94}, 769-786.

\bibitem[Lijoi et al.(2007a)]{Lij(07a)}   
Lijoi, A., Mena, R.H., and Pr\"unster, I. (2007a). A Bayesian nonparametric method for prediction in EST analysis. \textit{BMC Bioinformatics}. \textbf{8}: 339.

\bibitem[Lijoi et al.(2007b)]{Lij(07b)}   
Lijoi, A., Mena, R.H., and Pr\"unster, I. (2007b). Controlling the reinforcement in Bayesian nonparametric mixture models. \textit{J. Roy. Stat. Soc. B}. \textbf{69}, 769--786.

\bibitem[Lijoi and Pr\"unster (2003)]{Lij(03)}   
Lijoi, A. and Pr\"unster, I. (2003). On a normalized random measure with independent increments relevant to Bayesian nonparametric inference. \textit{Proceedings of the 13th European Young Statisticians Meeting}. Bernoulli Society, 123--134.

\bibitem[Mao(2004)]{Mao(04)}
Mao, C. X. (2004). Predicting the conditional probability of discovering a new class. \textit{J. Amer. Statist. Assoc.}. \textbf{99}, 1108--1118.

\bibitem[Mao and Lindsay(2002)]{Mao(02)}   
Mao, C.X. and Lindsay, B.G. (2002). A Poisson model for the coverage problem with a genomic application. \textit{Biometrika}. \textbf{89}, 669-681.

\bibitem[Navarrete et al.(2008)]{Nav(08)} 
Navarrete, C., Quintana, F., and M\"uller, P. (2008). Some issues on nonparametric Bayesian modeling using species sampling models. \textit{Stat. Model.}. \textbf{8}, 3--21.

\bibitem[Pitman(1995)]{Pit(95)}
Pitman, J. (1995). Exchangeable and partially exchangeable random partitions. \textit{Probab. Theory Related Fields}. \textbf{102}, 145--158.

\bibitem[Pitman(2003)]{Pit(03)}
Pitman, J. (2003). Poisson-Kingman partitions. In \textit{Science and Statistics: A Festschrift for Terry Speed} (Goldstein, D.R. Eds). Lecture notes monograph
series. \textbf{40}, Institute of Mathematical Statistics.

\bibitem[Pitman(2006)]{Pit(06)} 
Pitman, J. (2006). \textit{Combinatorial Stochastic Processes.} Ecole d'Et\'e de Probabilit\'es de Saint-Flour XXXII. Lecture Notes in Mathematics N. 1875. New York: Springer.

\bibitem[Pitman and Yor(1997)]{Pit(97)}
Pitman, J. and Yor, M. (1997). The two-parameter Poisson--Dirichlet distribution derived from a stable subordinator. \textit{Ann. Probab.}. \textbf{27}, 1870--1902.

\bibitem[Provost(2005)]{Pro(05)}
Provost, S.B. (2005). Moment-based density approximants. \textit{Math. J.}. \textbf{9}, 727--756.

\bibitem[Pr{\"u}nster, 2002]{prunster2002random}
Pr{\"u}nster, I. (2002).
\newblock {Random probability measures derived from increasing additive
  processes and their application to Bayesian statistics}.
\newblock {\em Ph. D. Thesis, University of Pavia}.

\bibitem[Rasmussen and Starr(1979)]{Ras(79)} 
Rasmussen, S.L. and Starr, N (1979). Optimal and adaptive stopping in the search for new species. \textit{J. Amer. Statist. Assoc.}. \textbf{74}, 661--667.

\bibitem[Regazzini et~al., 2003]{regazzini2003distributional}
Regazzini, E., Lijoi, A., and Pr{\"u}nster, I. (2003).
\newblock Distributional results for means of normalized random measures with
  independent increments.
\newblock {\em Annals of Statistics}. 560--585.

\bibitem[Ruggiero et al.(2015)]{Rug(13)} 
Ruggiero, M., Walker, S.G., and Favaro, S. (2013). Alpha-diversity processes and normalized inverse Gaussian diffusions. \textit{Ann. Appl. Probab.}. \textbf{23}, 386--425.

\bibitem[Sampson(2001)]{Sam(01)} 
Sampson, G. (2001). \textit{Empirical Linguistics.} Continuum, London - New York.

\bibitem[Steinbrecher and Shaw(2008)]{Ste(08)} 
Steinbrecher, G. and Shaw, W.T. (2008). Quantile mechanics. \textit{European J. Appl. Math.}. \textbf{19}, 87--112.

\bibitem[Susko and Roger(2004)]{Sus(04)}
Susko, E. and Roger, A. J. (2004). Estimating and comparing the rates of gene discovery and expressed sequence tag (EST) frequencies in EST surveys. \emph{Bioinformatics}. \textbf{20}, 2279--2287.

\bibitem[Zhang(2005)]{Zha(05)}
Zhang, C.H. (2005). Estimation of sums of random variables: examples and information bounds. \textit{Ann. Statist.}. \textbf{33}, 2022--2041.

\end{thebibliography}
\end{document}